\documentstyle[twocolumn,aps]{revtex}

\begin{document}
\draft
\preprint{}
\title{Electromagnetic forces in photonic crystals}
\author{M. I. Antonoyiannakis\cite{byline} and J. B. Pendry}
\address{
      Condensed Matter Theory Group, The Blackett Laboratory, 
              Imperial College, London SW7 2BZ, UK}

\date{\today}
\maketitle
\begin{abstract}
We have developed a general methodology for numerical computations of electromagnetic (EM) fields and forces in matter, based on solving the macroscopic Maxwell's equations in real space and adopting the Maxwell Stress Tensor formalism. Our approach can be applied to both dielectric and metallic systems of frequency-dependent dielectric function; as well as to objects of any size and geometrical properties in principle. We are particularly interested in calculating forces on nanostructures.

Our findings confirm that a body reacts to the EM field by minimising its energy, {\it i.e.} it is attracted (repelled) by regions of lower (higher) EM energy. When travelling waves (of real wavevector) are involved, forces can be additionally understood in terms of momentum exchange between the body and its environment. However when evanescent waves (of complex wavevector) dominate, the forces are complicated, often become attractive and cannot be explained by means of real momentum being exchanged. 

We have studied the EM forces induced by a laser beam on a crystal of dielectric spheres of GaP. We observe effects due to the lattice structure, as well as due to the single scattering from each sphere. In the former case the two main features are a maximum momentum exchange (and largest forces) when the frequency lies within a band gap; and a multitude of force orientations when the Bragg conditions for multiple outgoing waves are met. In the latter case the radiation couples to the EM eigenmodes of isolated spheres (Mie resonances) and very sharp attractive and repulsive forces occur. Depending on the intensity of the incident radiation these forces can overcome all other interactions present (gravitational, thermal and Van der Waals) and may provide the main mechanism for formation of stable structures in colloidal systems. 
\end{abstract}

\narrowtext
 
\section{Introduction}
\label{sec:Introduction}

\subsection{Photonic structures}  
\label{subsec:general intro}

Photonic materials possess structure (crystalline order) at the scale of the wavelength of light which they therefore diffract strongly, just like x-rays are diffracted from atomic crystals. By adjusting the geometric and dielectric parametres of a photonic crystal, we can alter the {\it optical} properties at will\cite{PB.1,PB.2,PB.3,PB.4,PB.5}, in the same way as a semiconductor crystal can be built to support desirable {\it electronic} properties. For example, the dispersion relation (band structure) of light inside a photonic crystal is very different from the linear $ \omega = k \,c$ behaviour we have in homogeneous media. Another startling effect is the possibility for no allowed optical modes to exist at certain frequency ranges (band gaps) which, coupled with the ability to inject optical modes at specific frequencies by introducing defects in the crystalline structure, has fuelled prospects for novel optical phenomena such as photon localisation, inhibition of spontaneous emission and enhanced nonlinear effects. Novel devices (single-mode LEDs, thresholdless lasers, etc.) are also expected. 

Where do we find photonic crystals ? Natural gemstone opals owe their beautiful iridescent colours to their structure which strongly diffracts visible light. However their relatively large amount of disorder as well as the poor refractive index contrast they display
renders them impractical for the detailed tailoring of optical properties one needs to attain if the above mentioned novel effects are to be sought after. The first man-made photonic crystals were made ten years ago by Yablonovitch and they operated at the scale of microwaves. Intensified efforts for the past years have born fruit and we are now starting to have artificial opals at visible wavelengths, fabricated by a wealth of experimental techniques which bears witness to the ingenuity with which the problem has been tackled. Some methods aim at building crystals from colloidal suspensions utilising the influence of gravitational, electromagnetic or biochemical interactions, while others concentrate on selectively removing material from a matrix (etching) and thus forming a desired structure. `Mixed' approaches are also used (removing material followed by ordering or stacking of crystal layers)\cite{Fab}.

The theoretical understanding of photonic structures has centred at solving Maxwell's equations for the EM wavefield in the presence of matter. Thus the classical theory of vector fields is applied. This methodology, appropriate for scales large enough compared to the underlying atomic structure, has been successfully applied to predict both the dispersion relation (band structure) as well as the scattering properties of such systems. 

\subsection{Overview of photonic forces}  
\label{subsec:overview}

Nature favours phenomena that are energetically profitable. For example, the Van der Waals force between two macroscopic objects can be found by calculating the change in the self-energy of the vacuum-fluctuation wavefields which multiply scatter off the objects. In electronic band structure, total energy calculations combined with variational arguments have been successfully used to predict the equilibrium structure of crystals and to realistically estimate a plethora of physical quantities\cite{EBS-force}. But while the former example involves virtual light waves and the latter employs real electronic waves, there is an increasingly widening range of systems where real photons apply; of those, photonic crystals form a subset whose potential for new physics and new applications is only beginning to be realised.

In such a crystal, the photonic band structure contains the energy levels accessible to photons. But unlike atomic crystals, where the sea of electrons occupies a range of modes, it is possible with an external source of monochromatic light to selectively populate a particular mode in a photonic crystal. The light source injects many photons but owing to the bosonic and non-interacting nature of light, each photon would remain in the same state. Thus photonic crystals are suitable for observing single-mode band-structure effects, a property implying rich force spectra. To give an example, consider what happens as we traverse with a laser of tunable frequency the first band gap in a photonic crystal made of dielectric spheres in air. On the low energy side, the EM wavefield is concentrated outside the spheres ({\it i.e.} at regions of low dielectric constant), hence the energy of this mode is lower. Inside the gap, the fields are everywhere zero since all light is expelled from the crystal. And on the high energy side, the field is concentrated inside the spheres. This shifting of the wavefield from one region to another translates into changes in the spatial distribution of EM energy inside the crystal, and would therefore strongly affect the forces. 

Apart from performing total energy calculations, EM forces can be also computed in an alternative, more formal manner: namely within the Maxwell Stress Tensor (MST) formalism. In this methodology, the various components of the MST are constructed from the electric and magnetic fields in real space. Integration of the tensor around a closed surface surrounding the body of interest yields the total force and torque acting on it. There is in principle no restriction as to the size and shape of the body, nor as to its dielectric properties. The generality of this method makes its applicability wide, ranging from purely electric to purely magnetic to fully electromagnetic effects; from virtual to real photons; and from travelling to evanescent wavefields.

Our approach in this paper is as follows: First we explain in some detail how the concept of a stress tensor arises in the calculation of forces. We define the tensor for the case of electromagnetism and show how it can be incorporated in numerical calculations of photonic effects. We then put our methodology at work. As a demonstration of some principal features of EM forces we treat first the case of a beam of laser light scattering from an homogeneous medium (metallic or dielectric) of variable thickness. This is followed by the main body of our results which regards light-induced forces on dielectric crystals. 

\section{Methodology}
\label{sec:Methodology}

\subsection{Theoretical framework}  
\label{subsec:theo_frame}
\subsubsection{The need for a stress tensor}  
\label{subsubsec:need_tensor}
 
We are interested in calculating EM forces on macroscopic bodies in the nanometre scales. It is therefore natural to seek expressions for the forces in terms of macroscopic quantities (such as dielectric functions). We therefore make at the outset an implicit adoption of the continuum picture of matter. We follow the treatment of Landau and Lifshitz\cite{LL.1}.

A force acting on a macroscopic body is a sum of the forces ${\bf f}_v({\bf r})$ on each of the volume elements that constitute the body. Thus the total force ${\bf F}$ is
				\begin{equation} 
{\bf F} = \int {\bf f}_v({\bf r}) \,dv \,. \label{eq:fdV}
                            	\end{equation}
If ${\bf f}_v$ can be expressed as the divergence of a tensor of rank two, then we can transform the volume integral into a surface integral and the total force becomes
				\begin{eqnarray} 
 F_\alpha &=& \int \frac{\partial {\tensor T}_{\alpha \beta}}{\partial x_\beta} \,dv \,,  \;\;\;\;\;\;\;\; 
\alpha, \beta = \{x,y,z\} \,, \label{eq:T'dV} \\
&=& \oint {\tensor T}_{\alpha \beta} \,dS_\beta \;, \label{eq:TdS}
                            	\end{eqnarray}
where a sum over repeated indices is implied from here onwards. The {\it stress tensor} ${\tensor T}_{\alpha \beta}$ contains all the information we need. 
Note it has dimensions of energy density. This is not accidental, since forces can be also thought to arise due to spatial variations in the energy (energy-gradient picture). They act in order to lower the system's energy {\it i.e.} to push the system into regions where the energy is lowest. Therefore only the spatially varying part of ${\tensor T}_{\alpha \beta}$ contributes to forces, a fact which can be seen in Eq.\ (\ref{eq:T'dV}). Eq.\ (\ref{eq:TdS}) is of great practical benefit in that it is often easier to calculate the stress tensor on the surface of a solid body rather than in its interior.

\subsubsection{The stress tensor for electromagnetism}  
\label{subsubsec:EMtensor}

So far the discussion is general. To concentrate on EM forces we need the stress tensor for the case of electromagnetism. When a solid body is immersed in a linear, isotropic medium of permittivity $\epsilon_0 \epsilon_m$ and permeability $\mu_0 \mu_m$, then the spatially non-constant part of the stress tensor in the region external to the body is given in terms of the electric (${\bf E}$) and magnetic (${\bf H}$) fields by \cite{LL.2} 

				\begin{eqnarray} 
{\tensor T}_{\alpha \beta} &=& \epsilon_0 \epsilon_m E_{\alpha} \,E_{\beta} + \mu_0 \mu_m H_{\alpha} \,H_{\beta} \nonumber \\
&& \;\; - \frac{1}{2} \,\delta_{\alpha \beta} \,( \epsilon_0 \epsilon_m E_{\gamma} \,E_{\gamma} +
\mu_0 \mu_m H_{\gamma} \,H_{\gamma} )  \label{eq:MSTdef} \,
				 \end{eqnarray}
provided the medium remains in mechanical and thermal equilibrium under the influence of the EM fields ({\it i.e.} electrostriction and magnetostriction effects are ignored). Thus by integrating ${\tensor T}_{\alpha \beta}$ of Eq.\ (\ref{eq:MSTdef}) over a closed surface ${\bf S}$ enclosing the body, the {\it full} EM force on the body is obtained. No further  approximations or limiting assumptions need be made, as has often been done with other methodologies employed for EM effects (e.g. dipole approximation, vanishing particle size compared to the wavelength of light in scattering phenomena, etc.). Our methodology can be applied to dielectric or metallic objects of any shape and at any frequency range in principle. The only requirement for meaningful results is that the integration is carried out over scales which are large with respect to atomic sizes, so that the continuum picture of matter is justified. For nanostructure scales of interest to us ($> 50$ nm) this requirement is surely satisfied: the dynamics of the EM wavefield is more properly described by the equations of Maxwell rather than by the Schr{\"o}dinger equation for all the electronic charges which make up the dielectric response of the system. Finally, our choice of macroscopic scales dictates a relevant timescale also (no macroscopic object can respond synchronously to THz oscillations): we must time-average the forces ({\it i.e.} the stress tensor) over many cycles when dealing with harmonic fields at optical frequencies.

\subsection{Numerical methods}
\label{subsec:num_meth}

We discretise Maxwell's equations on a simple cubic (SC) mesh. For calculations of photonic band structure, periodic boundary conditions are used in all directions. For transmission results, we consider a crystal finite in the $z$ direction, having a thickness of $N$ unit cells (layers), $N$ being allowed to vary. We shall discuss here the case where a monochromatic light beam of frequency $\omega$, polarisation $\sigma$ and wavevector ${\bf k}$ is incident on the crystal from the $-z$ side. Using the Transfer Matrix formalism we calculate reflection coefficients 
$R_{\bbox{k'}\sigma'}$ for scattering into plane waves of wavevector 
${\bf k'} = {\bf k_\parallel}+{\bf g_\parallel} + {\bf\hat{z}} k'_z$ and polarisation $\sigma'$, where ${\bf k_\parallel}$ is the component of ${\bf k}$ parallel to the $xy$ plane and ${\bf g_\parallel}$ is a {\it two-dimensional} reciprocal lattice vector also lying on the $xy$ plane. Our computer codes have been published and issues of stability and accuracy have been discussed elsewhere \cite{OPAL}. 

The new ingredient is the calculation of forces. From the reflection coefficients we construct the total fields on the $-z$ side of the crystal,
				\begin{mathletters} 
\label{eq:field_reconstruction}	
			        \begin{equation} 
{\bf E}({\bf r},t) = 
{\bf E}_{inc}^{0} 
\sum_{\bbox{k'},\sigma'} 
( \delta_{ \bbox{k},\bbox{k'} }  \delta_{ \sigma,\sigma' } +
  R_{\bbox{k'}\sigma'} ) \;
{\bf \hat{e}}_{\bbox{k'}\sigma'} \;
e^{i [{\bf k'} \cdot {\bf r} - \omega t ]} \,,
 \label{eq:fieldE} 
			        \end{equation}
			        \begin{equation} 
{\bf H}({\bf r},t) = 
{\bf H}_{inc}^{0} 
\sum_{\bbox{k'},\sigma'} 
( \delta_{ \bbox{k},\bbox{k'} } \ \delta_{ \sigma,\sigma' } +
  R_{\bbox{k'}\sigma'} )\;
{\bf \hat{e}}_{\bbox{k'}\sigma'} \;
e^{i [{\bf k'} \cdot {\bf r} - \omega t ]} 
 \label{eq:fieldH} \,.
			        \end{equation}
				\end{mathletters} 
In Eq.\ (\ref{eq:field_reconstruction}) ${\bf E}_{inc}^{0},{\bf H}_{inc}^{0}$ are the amplitudes of the incident fields and ${\bf \hat{e}}_{\bbox{k'}\sigma'}$ are the unit vectors for each ${\bf k'}$ and each polarisation $\sigma'$. The Kronecker delta functions serve as to add the contribution from the incident field, which is present on the $-z$ side, to the reflected field. For the total fields on the $+z$ side the only contribution comes from the transmitted fields. 

Having constructed the total fields outside the crystal, we invoke once again the Matrix Transfer equations to find the total field at each mesh point inside the structure. When this is accomplished, the stress tensor components are calculated and time-averaged. The latter procedure can be made in a standard way since the fields are expressed as sums of plane waves at the same frequency $\omega$. Finally, the forces are found by integrating the stress tensor over a surface enclosing the body of interest, which can be of arbitrary shape and complexity. In this work we present force calculations on isolated spheres, crystal layers and the entire crystal sample.

\section{Results and Discussion}
\label{sec:Results}

\subsection{Homogeneous system - travelling waves}  
\label{subsec:Homog_system_trav}

Probably the simplest system for studying macroscopic radiation forces on materials is a homogeneous half-space (dielectric or metallic) upon which a single beam of light is incident. The simplicity of this model system will allow us to put forward in a transparent manner some important features of our methodology. Furthermore, as this problem is analytically solvable it provides a nice test-ground for our numerical results. 

We model the semi-infinite medium by considering a layer of material of sufficient thickness in the $z$ direction so that all light entering the layer is absorbed before reaching the far end. For a metal this would only take a thickness equal to the skin depth (a few nm typically) when $\omega$ lies below the plasma frequency $\omega_p$; for a real dielectric, the tiny absorption present makes it necessary to consider much thicker samples, perhaps as much as several cm. In either case, only the incident and specularly reflected wavefields are present outside the system. The force on the layer is found by integrating the stress tensor for the total field (incident plus reflected) on a surface $S_z$ which lies on the $-z$ side immediately above the medium and extends to infinity in the $xy$ plane (the contribution from the integration on the far end is zero, since the fields vanish there). A more appropriate quantity for this system is the EM pressure $P_\alpha = (1/S_z) \oint {\tensor T}_{\alpha z} dS_z$ on the layer.  

As a dielectric system we chose GaP which is transparent ({\it i.e.} has low absorption and the dielectric function $\epsilon$ is mostly real and positive) over near infrared and visible frequencies but behaves as a poor metal ({\it i.e.} there is a plasma frequency at $\sim$ 13.5 eV but the absorptive part of $\epsilon$ becomes significant with respect to the real part) in the ultraviolet spectrum. As a metallic system we use Al, for which the dielectric function can be modelled as 
			        \begin{equation}
\epsilon = 1 - \frac{\omega_p^2}{\omega(\omega + i \gamma)} \,, 
 \label{eq:eps_Al}	
        		        \end{equation}
where $\hbar \omega_p=15$ eV and the damping coefficient for the intraband transitions is $\hbar \gamma = 0.1$ eV. Our data for the dielectric properties of these materials come from \cite{eps_Al}.

We summarise our main findings for the forces on this system:
\\
(a) A homogeneous system can only scatter light into directions belonging 
to the scattering plane (the plane defined by ${\bf k}$ and ${\bf \hat z}$) so EM forces must lie on this plane. Also the polarisation of light is conserved upon scattering. 
\\ 
(b) The response to light incident on a homogeneous half-space will range from full reflection (e.g. for a perfect metal at frequencies below $\omega_p$) to full absorption (for a dielectric at the Brewster angle and $p$-polarised light). For normal incidence, the pressure on the half-space will range respectively from $2 P_{inc}$ (full reflection) to $P_{inc}$ (full absorption), where  $P_{inc} = I_0/c_0$ is the radiation pressure of the incident beam which has intensity $I_0$ and travels at the free-space speed $c_0$. If we replace the half-space with a layer of finite thickness such that some light is transmitted through the far end, taking care to include the contribution from the far end in the stress tensor integration, then the lower bound for the force becomes zero, corresponding to full transmission ({it i.e.} zero reflection and zero absorption). 
These effects are demonstrated for GaP and Al samples in Figs.~\ref{fig.1}, \ref{fig.2}. It is clear from these plots that the force on the system equals minus the net momentum change of the light, just as one would expect from the action-reaction principle. Thus the explanation of forces in terms of the momentum-balance picture is valid.  

The three values (0,$P_{inc}$ and $2 P_{inc}$) constitute the natural limits for {\it total} ({\it i.e.} external) EM forces on a macroscopic body (homogeneous or not, finite or not) of arbitrary size, due to incidence of a single beam of light. The only way for these limits to be exceeded is when a resonance condition is fulfilled. We shall come to this interesting case when dealing with forces on dielectric crystals. 
%
%
%
\noindent
(c) Since forces can just as well be attributed to spatial variations of energy as to temporal variations of momentum, we can use the energy-gradient picture to study EM forces on bodies. In this picture, the EM force on an infinitesimal volume element located at {\bf r} is 
			        \begin{equation}
{\bf F} = - \nabla U \,, 
 \label{eq:gradU}	
        		        \end{equation}
where $U$ is the energy of the EM wavefield at ${\bf r}$. The force on a macroscopic body is found by adding the contributions from all elements comprising the body. For example, when light is incident at normal angle on a layer of material extending from $0$ to $d$ in the $z$ direction, the total ({\it i.e.} average) force on the layer is 
			        \begin{equation}
{\bf F} = - \frac{1}{d} \int_0^d \nabla_z U \ dz 
= - \ {\bf \hat z} \ \frac{U(d)-U(0)}{d} \;. 
 \label{eq:gradUtot}	 
        		        \end{equation}
We show in Fig.\ \ref{fig.2} the total pressure on a layer of Al, 400 nm in thickness. The pressure is calculated both with the stress-tensor method (solid grey line) and with the energy-gradient method (dotted grey line). The agreement between the two is excellent, demonstrating the validity and usefulness of the energy-gradient picture. After all, the physical basis for forces in nature is the lowering of the energy, and electromagnetism is no exception to this. Thus the sharpness of the spatial variations of the stress tensor (or the energy) is a measure of how strong and localised forces are. Smooth, continuous variations of ${\tensor T}_{\alpha \beta}$ imply forces that vanish when the volume element they act on shrinks to zero size. Surface forces, in contrast, are localised at the interface between media; they are associated with abrupt, discontinuous changes in the stress tensor and can be explained physically as spatial discontinuities in the energy across the interface.

\subsection{Homogeneous system - evanescent waves}  
\label{subsec:Homog_system_evan}

The next level of complexity involves homogeneous systems which are exposed to evanescent wavefields. In such fields the light wavevector {\bf k} has got an imaginary component so that the dispersion relation is always satisfied:
			        \begin{equation}
\epsilon = {\bf k} \cdot {\bf k} \,, 
 \label{eq:e=k^2}	
        		        \end{equation}
where $\epsilon = \epsilon' + i \epsilon''$ is the dielectric function of the medium of propagation.
 
Our system is a dielectric layer (denoted as medium II) of thickness $d$, separating two dielectric half-spaces I, III (see Fig.\ \ref{fig.homog.layer}), with the planar interfaces lying in the $xy$ plane. For simplicity we choose all media to have zero absorption. A beam of light of wavelength $\lambda$ is incident from medium I at an angle $\theta_1$ and a wavevector 
${\bf k}_1 = {\bf {\hat x}} \beta + {\bf {\hat z}} \gamma_1 $. Upon penetration into region II, the wavevector may pick up an imaginary component along ${\bf {\hat z}}$:
			        \begin{eqnarray}
\epsilon'_2 &=& \beta^2  +  \gamma'^2 \;\;; \;\;\;\;
{\bf k}_2 = (\beta,0,\gamma'_2) \,,\;\;\;\text{ when}\; \beta^2 < \epsilon'_2 \\
\epsilon'_2 &=& \beta^2  -  \gamma''^2 \;\;; \;\;\;
{\bf k}_2 = (\beta,0, i \gamma''_2) \,,\;\;\;\;\text{ when}\; \beta^2 > \epsilon'_2 \;.
        		        \end{eqnarray}
By choosing $\epsilon'_2 < \epsilon'_1$ and adjusting $\theta_1$  (since $\beta = k_1 \cos\theta_1$) appropriately, we can create an evanescent wave in region II with the desirable imaginary momentum $\gamma''_2$. Here we concentrate on the case of medium II being vacuum ($\epsilon'_2 = 1$). We discuss {\it total} forces on media I and III, not interface forces.

When only travelling waves exist in all three media, the two half-spaces repel each other. In this case, the photons can be thought of as tennis balls: as they scatter off the walls of I and II, they exert a positive pressure on each wall. 

However, when ${\bf k}_2$ has a sufficiently large imaginary component along ${\bf {\hat z}}$, the force changes sign and becomes attractive. In this case the tennis-ball analogy breaks down. Our results indicate that at  the electrostatic limit $\lambda \gg d$, the total force on III per unit area varies as 
			        \begin{equation}
P_z = \epsilon_0 \frac{\gamma^2_1 (|\gamma_3|^2-\gamma''^2_2)}{|\gamma'_1 + \gamma_3|^2}  
\label{eq:P_z}	\;. 
                                \end{equation}
Thus the force becomes zero when $|\gamma_3|=\gamma''_2$. Curiously, this is the same condition for the angles of refraction $\theta_2, \theta_3$ to be equal, since
			        \begin{equation}
|\tan\theta_i|  = \frac{\beta}{|\gamma_i|} \,, \;\;\;\;\;\;\;\; 
 \label{eq:tantheta}	
i = \{1,2,3\} \,.     		
                                \end{equation}
When $\gamma''_2<|\gamma_3|$ the force is repelling, and 
$\theta_2 > \theta_3$, just as in the case of a travelling wave scattering off a denser medium. On the other hand, when $\gamma''_2 > |\gamma_3|$, we have $\theta_2 < \theta_3$, which is reminiscent of a travelling wave scattering off a dense/rare interface; but at the same time the total force is attractive, a fact for which there is no analogue in waves of real ${\bf k}$.

The magnitude of the attractive force is sensitive to the ratio of the wavelength $\lambda$ to the layer thickness $d$ over which the evanescent fields decay before entering medium III. We find that for $\lambda/d \sim 5$ the force has already dropped by more than three orders of magnitude compared to its value at the electrostatic limit $\lambda \gg d$. In Fig.\ \ref{fig.evan} we plot the normal pressure $P_z$ versus $\cos\theta_1$ for three cases, (i) $\epsilon'_1 < \epsilon'_3$, (ii) $\epsilon'_1 = \epsilon'_3$ and (iii) $\epsilon'_1 > \epsilon'_3$. For (i) and (ii) the attractive part of $P_z$ displays a smooth maximum, whereas for (iii) it peaks sharply. In all cases the highest value of $P_z$ at the electrostatic limit is of the same order of magnitude as the incident radiation pressure. For (i) and (ii) the field in region III is that of a travelling wave for all angles and therefore $|\gamma_3| > 0$. However for (iii) the field in III is evanescent when $\theta_1 > \theta_0$, $\theta_0$ being the angle for total internal reflection at an interface between media I and III\footnote{It is interesting to note that a sharp feature in $P_z$ is present always at $\theta_0$, provided  $\epsilon'_1 > \epsilon'_3$. Its position in the $\theta$ axis is independent of the dielectric properties of layer II.}. At $\theta_0$ we have $|\gamma_3| = 0$. From Eq.\ (\ref{eq:P_z}) it becomes clear why a vanishing $\gamma_3$ results into a sharp peak in $P_z$. For moderate absorption in III,  $|\gamma_3|$ does not quite reach zero at $\theta_0$ and the strength of the attraction is reduced although the peak is still clearly seen: for example, the maximum drops by 40 $\%$ when $\epsilon''_3 \sim 0.1$. 
Since the EM force is proportional to the light intensity $I_0$, we can adjust the latter so that the force induced by the evanescent (or the travelling) wavefield supersedes the dispersion forces between the half-spaces. We have indicated the strength of the Van der Waals attraction in the figure, calculated in the macroscopic Lifshitz theory with retardation and non-additivity effects considered. Owing to the $d^{-4}$ behaviour of the Van der Waals force\cite{VdW-ret}, only large enough separations will allow the photo-induced force to dominate.

%
%

\subsection{Dielectric crystals}  
\label{subsec:Diel_crystals}

We now turn our attention to a crystal of dielectric (GaP) spheres arranged in a SC lattice. GaP has an essentially real, constant dielectric function (modelled here as $\epsilon=8.9$) over the frequency range of relevance to this work (0.07-0.9 eV). The actual absorption of GaP is too small to have any sizeable effect on our force calculations, as we later show. We consider spheres with a radius $r = 200-400 $ nm, and choose a lattice spacing $l = 900$ nm, so the spherical surfaces are separated by $D=100-500$ nm. We often consider the reference medium in which the spheres are embedded to be air ($\epsilon_{med}=1$), although, for reasons of experimental feasibility, it is really a liquid medium we have in mind. Such a medium does not alter qualitatively the effects we describe provided its absorption is low enough. For example, we show that water is suitable. Most of our results are for a discretisation mesh $10\times10\times10$. The crystal is infinite in the $x$ and $y$ directions, and has thickness $N$ unit cells in the $z$ direction.

For the EM properties of this crystal we expect the following behaviour: 

\noindent
(a) Propagation of EM waves in an ``effective medium'' at large wavelengths $\lambda \gg l$ where light senses the crystal as an homogeneous dielectric medium. 

\noindent
(b) Multiple scattering effects: 
{\it i)} appearance of forbidden ranges of frequencies as $\lambda \sim l$ where the multiply scattered waves interfere destructively and annihilate each other inside the crystal; 
{\it ii)} scattering by specific crystal planes into non-specular directions as $\lambda^{-1} \sim |{\bf k} + {\bf G}|/(2 \pi)$, ${\bf G}$ being a reciprocal lattice vector. 

\noindent
(c) Resonant features at frequencies corresponding to the EM eigenmodes of each sphere.

We observe these effects in Fig.\ \ref{fig.pbs.T.force}, where the band structure (dispersion relation) is shown alongside transmittance and force for one layer of thickness, normal incidence. At low energies $\omega$ is proportional to the wavevector $k_z$, just as in free-wave propagation, the constant of proportionality being the wave velocity of a homogeneous effective medium replacing the crystal. For a system of unconnected spheres like ours, the dielectric function $\epsilon_{eff}$ of the effective medium is well approximated within the Garnett theory\cite{Garnett},
			        \begin{equation}
\epsilon_{eff} = \frac{\epsilon_m (1 + 2 A)}{1-A}\,, \;\;\;\;\;\;\;\; 
A=f \frac{\epsilon - \epsilon_m}{\epsilon+ 2 \epsilon_m} \,, 
 \label{eq:e_eff}	 
        		        \end{equation}
$f$ being the volume fraction of the spheres in the crystal. For our system Eq.\ (\ref{eq:e_eff}) yields $\epsilon_{eff} = 1.76$. We find that at low energies (or wavelengths $\lambda > 10 \cdot l$) the force on the crystal is well approximated by the analytical expression for the force on a homogeneous film of the same thickness and dielectric function $\epsilon_{eff}$. 
%
%

At higher energies band gaps open up and the propagation of all EM modes is forbidden inside an infinite crystal: all incident light will be reflected back when the frequency lies within a band gap. It is impressive that even for one layer of thickness the major band gap centred at around 0.7 eV causes a low transmission. Consequently the pressure on the layer approaches the upper natural limit $2 P_{inc}$ for frequencies inside this gap. The lowest band gap requires a thicker sample before it can clearly show its signature in transmission and force spectra, but otherwise it affects the system's behaviour in the same way. 

Another consequence of multiple scattering is that under certain conditions, the scattered fields can interfere constructively and produce outgoing waves in additional directions to those of the specularly reflected and directly transmitted beams. We call this scattering {\it off-diagonal}, since when it occurs the wavevector of the scattered field picks up a 2D reciprocal lattice vector component,
			        \begin{equation}
{\bf k'} = {\bf k_\parallel}+{\bf g_\parallel} + {\bf\hat{z}} k'_z \;.
 \label{eq:k'}	 
          		        \end{equation}
Most often the energetics of the problem are such that $k'_z$ is imaginary and all the off-diagonal fields are evanescent. However at certain ranges of frequencies a real $k'_z$ is allowed and the off-diagonally scattered beams are travelling waves. For a SC lattice the minimum energy for this to occur is when ${\bf k_\parallel}$ lies at the edge of the first Brillouin zone along the $\Gamma X$ direction. Considering an incident beam with ${\bf k}=(k_x,0,k_z)$, the condition becomes 
			        \begin{equation}
k \sin\theta = \frac{\pi}{l} \;\Rightarrow\; \omega = \frac{\pi}{l} \, \frac{c_0}{\sin\theta} \;.
 \label{eq:ODS}	 
          		        \end{equation}
Eq.\ (\ref{eq:ODS}) gives the lowest frequency for off-diagonal scattering for a crystal of lattice constant $l$ and a wave incident at angle $\theta$. 

As one imagines, the effects on forces are rich as the possible combinations of momentum-carrying outgoing waves are endless. To give one example, consider the incidence at $\theta=70 \deg$ of a laser beam on our crystal of one layer thickness. For this system we plot the intensity of reflection $r$ and transmission $t$ of each outgoing wave in Fig.\ \ref{fig.ODS.R}. Off-diagonal scattering sets in for $g_\parallel = (-2 \pi/l,0,0)$ at frequency 0.71 eV in good agreement with Eq.\ (\ref{eq:ODS}). At slightly higher frequencies the extra energy is given to $k'_z$ while $k'_x$ is fixed to the value $\pi/l$, provided the angle of incidence is not changed. Thus the outgoing waves include two reflected and two transmitted beams, all forming an angle $\theta$ with the surface normal (Fig.\ \ref{fig.ODS.beams}). In addition, notice that at 0.76 eV the intensities of the leftwards and the rightwards travelling waves are equal: $r_{\rightarrow} + t_{\rightarrow} = r_{\leftarrow} + t_{\leftarrow}$. This means that the total momentum of the scattered EM fields along ${\bf {\hat x}}$ is zero. In other words, the crystal absorbs all incident momentum along ${\bf {\hat x}}$, and the $x$-force on a surface $S$ equals 
  			        \begin{equation}
F_x = (P_{inc} \,S) \sin\theta \,.
 \label{eq:F_x}	 
          		        \end{equation}
Our force calculations  confirm that $F_x$ approaches this value at $\omega \sim 0.76$ eV.
%
%

Off-diagonal scattering survives a moderate amount of absorption in the crystal. We find that for $\epsilon''=0.1$ the reflectivities of the off-diagonal beams decrease by less than 40\% (depending on the frequency), which is a little larger than the decrease the diagonal beams suffer ($<20\%$).

Thirdly, we note that at specific frequencies there appear highly flat modes in the band structure combined with sharp features in the transmittance spectrum (for example at 0.624,0.768,0.776,0.786,0.85,0.88 eV on the $k_z = 0$ axis, Fig.\ \ref{fig.pbs.T.force}). Due to their low dispersion they possess a low group velocity $d \omega/d k$ and for this reason they have been conveniently termed `heavy photons' by Ohtaka and Tanabe\cite{Ohtaka.heavy}. As has been argued elsewhere\cite{Ohtaka.heavy,p1}, these are resonance peaks corresponding to the EM eigenmodes of isolated spheres ({\it Mie resonances}). Their presence strongly affects the position of the band gaps on the frequency axis\cite{PB-Mie}. The eigenmodes are spherical waves labelled by three quantum numbers, just as in the case of electronic orbitals in atoms, and are generally electric or magnetic oscillating multipoles. Due to the spherical symmetry they possess a $(2 n+1)$ degeneracy, $n$ being the azimuthal quantum number. However the presence of a plane wave incident along the direction of ${\bf k}$ breaks this symmetry and modifies the boundary conditions so that {\it only one} eigenmode for each $n$ can be excited; this, at any rate, is the standard Mie theory result and strictly applies to the case of an isolated sphere.
The notation for the excitable electric (magnetic) modes is $a^i_n$ ($b^i_n$), $i$ being the sequential index of the resonance in the frequency axis. For dielectric spheres the electric modes $a_n^i$ are broader whereas the magnetic modes $b^i_n$ are sharper, the sharpness of both types of modes increasing with frequency. For metallic spheres, only the electric modes survive, being better known to some as `surface plasmons'. The first few Mie resonances for one GaP sphere subjected to a plane wave are shown in Table \ref{table1}. 
%
%

When many spheres approach each other to form the lattice, interactions occur which shift the resonances with respect to the frequencies for an isolated sphere (Fig.\ \ref{fig.w.r}). Unless the coupling between neighbouring spheres is strong, each resonance in the crystal will retain its characteristic EM field distributions on the surface and the interior of each sphere. Thus the lowest-in-frequency heavy-photon resonance in our GaP crystal is the $b_1^1$ magnetic dipole mode, followed by the electric dipole $a_1^1$ mode (Fig.\ \ref{fig.modes}). The EM fields for these two resonances clearly match the expected patterns for Mie scattering from a single sphere. For the next two sharp peaks in transmittance the EM fields are similar but not identical to those for the magnetic quadrupole mode $b_2^1$; and likewise for the following two  
peaks, whereby the fields display partial similarity with those for the electric quadrupole mode $a_2^1$. These are cases of mixing from more than one Mie modes owing to the organisation of the spheres into a lattice. To quantify the contribution of each Mie mode into the heavy-photon resonance one has to decompose the scattered field into its constitutive spherical waves, as Ohtaka and Tanabe have done\cite{Ohtaka.decomp}. However the mode that follows in frequency is a distinct electric quadrupole, with field distributions identical to the $a_2^1$ Mie mode. 

Depending on $\omega$, the incident field produces polarisation charges and/or induced currents in the spheres and orchestrates their response according to which mode is excited each time. It is interesting that spheres which are initially non-magnetic ($\mu=1$) can acquire magnetic properties at the occurrence of any $b_n^i$ resonance. An effective homogeneous medium replacing the crystal structure would have $\mu\ne 1$ at this instance. 
%
%

What happens at a resonance? Each sphere acts as a cavity that traps the EM waves. The energy density internal to the spheres is 2-3 orders of magnitude larger than for the incident light beam\cite{p1,Mie_energy}. As Van Albada {\it et al} have shown\cite{v_E.1}, the transport of EM energy can be slowed down considerably at the conditions for Mie scattering. They derive a rigorous microscopic formula for the velocity of energy transport $v_E$, and subsequently propose a heuristic but physically plausible model for $v_E$, which they show to agree with the exact model and experimental results within reasonable accuracy, and which reads
  			        \begin{equation}
v_E = \frac{c_0}{f (W-1) + 1} \,,
 \label{eq:v_E}	 
          		        \end{equation}
where $c_0$ the speed of light in vacuum, $f$ the volume fraction and $W$ the energy density internal to the spheres relative to the surrounding medium (vacuum in our case). Our computations suggest that $W \sim 20$ for the first few heavy-photon modes, resulting in a velocity $v_E$ that is much lower than one expects from a straightforward estimate for the average velocity away from resonances,
 			        \begin{equation}
v_{av} = \frac{c_0}{f (\sqrt{\epsilon} - 1) + 1}    \;.      		         \label{eq:v_est}	 
                                \end{equation}
Thus for $f=0.22$ the equations above give $v_E/c_0 \sim 0.19$, whereas $v_{av}/c_0 \sim 0.69$. For comparison, the group velocity $v_g=d\omega/dk$ varies with frequency and ranges from $0.02 \;c_0$ to $0.25 \;c_0$ for these resonances. 
The accumulation of energy in the system at the resonance condition means that the overall EM fields are enhanced\footnote{This mechanism has been known to explain the enhancement of the field on the surface of GaP particles used in SERS (Surface Enhanced Raman Scattering) experiments \cite{Raman}.}, and this results in large forces when the stress-tensor integration is carried out.

The magnitude of the photo-induced force is proportional to $I_0$, the intensity of the incident beam. Provided the spheres are in a liquid medium which facilitates heat evacuation (so that melting is avoided), the photo-induced force can dominate gravitational, thermal-fluctuation and Van Der Waals forces. Assuming a moderate laser of $I_0 \sim 3.5 \,\times\,10^8$ W/m$^2$, the incident radiation pressure is $\sim 1$ pN per unit cell ($900\times900$ nm$^2$). For a crystal of one layer thickness, and when $\omega$ lies inside the second band gap (centred at 0.7 eV in Fig.\ \ref{fig.pbs.T.force}) the force on a unit cell is $\sim 2$ pN (the `natural limit' for full reflection), whereas at resonance conditions it exceeds this limit and can reach much higher values (e.g. 5 pN for the $b_1^1$ resonance). By contrast, the gravitational force on a GaP sphere is $m g \sim 0.8 \,\times\,10^{-3}$ pN. Thermal-fluctuation forces are comparable to gravity. Finally, the Van der Waals attraction between two GaP spheres\footnote{This value is obtained in the $D\ll r$ limit and ignoring retardation effects and is therefore an overestimate.} is less than 0.2 pN\cite{p1}.

The sharp force features shift slightly with the angle of incidence. More importantly, they are sensitive to the amount of absorption present in the spheres or the surrounding medium: a moderate $\epsilon''=0.1$ is enough to suppress them. For GaP spheres in water, however, we have found that the absorption is too small to blur the sharp features in forces, within the frequency range of interest (Fig.\ \ref{fig.f_vs_abs}). 
%
%

The effect of increasing thickness is of particular importance with reference to resonances. When the number of layers is doubled, the number of sharp features doubles also\cite{p1}. This effect is directly equivalent to degeneracy splitting in atoms. Two oscillating multipoles (GaP spheres in a heavy photon mode) brought close together interact, their fields hybridise and the two new energy levels of the combined system lie slightly below (at $\omega_<$) and slightly above (at $\omega_>$) the frequency of the original level (at $\omega_0$). This atomic physics analogy is far reaching: the mode at $\omega_<$ is bonding (attractive forces) whereas at $\omega_>$ it is anti-bonding (repulsive forces). We see a series of alternating signs in the force on a sphere as different resonances are encountered (Fig.\ \ref{fig.f_J=1}) for a crystal two layers thick. For the magnetic dipole ($b_1^1$) mode in particular, we have shown \cite{p1} that at $\omega_<$ all spheres have their dipoles parallel, whereas at $\omega_>$ the dipoles of the top layer are anti-parallel to those of the bottom layer. The same is true for the electric dipole ($a_1^1$) mode: all dipoles are parallel at $\omega=0.7915$ eV, anti-parallel at $\omega=0.8025$ for the system of Fig.\ \ref{fig.f_J=1}. Similarly, the multipoles of the higher-energy modes can be oriented constructively, in which case an attraction occurs, or destructively, leading to repulsion between neighbouring spheres. This argument holds also for the mixed hybrid states that are made out of more than one Mie modes, for example at 0.86 eV and 0.875 eV in Fig.\ \ref{fig.f_J=1}.

As expected, the splitting of the energy levels and the force strength increase as the spheres of two vertically adjacent layers approach each other (Fig.\ \ref{fig.f_split.with.dist}). 
The nature of the sharp forces between adjacent layers therefore lies in the sphere-sphere interactions {\it induced} by the light beam, when the latter populates the various Mie resonances. 

The analogy between resonant, light-induced, sphere-sphere interactions and the bonding of atoms that we just described is not the only one to be made. An analogy exists also with the case of cutting a metal in two along a plane, whereby the bulk plasmons reorganise into surface plasmons of attractive and repelling nature at frequencies below and above the original frequency\cite{Lucas}. The similarities between these three very different systems prompts us to acknowledge a general tendency in nature for the formation of energetically favourable EM bonds between two systems in the same state that come into proximity.

The rest of the force spectrum displays mostly smooth, attractive interactions due to the band gaps between the heavy-photon bands. This off-resonance attraction is produced as the band-gap condition for high reflectivity causes the top layer to absorb almost all the shock from the reflection of the incident light and be therefore pushed towards the bottom layer which remains relatively inert. 
%
%

Next we explore the possibility that the induced bonding interactions may lead to the formation of stable crystalline structures with full 3D periodicity. We start with two crystal layers on top of each other in a SC arrangement and gradually displace the top layer along the surface diagonal $(1,1,0)$, such that the relative distances between spheres of the {\it same} layer do not change, as shown in Fig.\ \ref{fig.diag.disp}. We monitor the normal ($F_z$) and shear ($F_x, F_y$) forces for the bonding mode of the magnetic-dipole $b_1^1$ resonance. 

We find that relative to a given sphere of the bottom layer ($A$), the shear force on the sphere above ($A_1$) alternates in sign as the top layer is displaced along the diagonal. For small displacement $\delta < 0.3$, the sphere $A_1$ is attracted back to its initial position $\delta_{init} = 0$ {\it i.e.} directly above $A$ of the bottom layer. As $\delta$ increases ($0.3 < \delta < 0.7$) the top sphere is attracted towards the void region in the centre of the square $ABCD$ ($\delta_{mid} = 0.5$), with a shear-force strength 5 times as large as the attraction to $\delta_{init}$. Finally for $0.7 < \delta < 1.0$ it is attracted towards $\delta_{fin} = 1.0$, {\it i.e.} directly above the sphere $C$ of the bottom layer, with a force strength similar to that for the attraction towards $\delta_{init}$. The normal force between the two layers remains attractive (since we are monitoring the bonding mode) but its magnitude displays a global maximum at $\delta_{mid}$ and two local maxima at $\delta_{init},\delta_{fin}$. Consequently, the spheres of the top layer experience two potential wells at $\delta_{init},\delta_{fin}$, whereby a stacking sequence $AAA...$ is obtained. However these wells are relatively shallow and when the potential barrier is crossed, the spheres find themselves in a much deeper well at $\delta_{mid}$ where a sequence $ABAB...$ is energetically favourable (Fig.\ \ref{fig.pot.wells}). As for the anti-bonding mode, there are two high potential barriers at $\delta_{init}$ and $\delta_{fin}$ and a barrier of lower height at $\delta_{mid}$. 
%
%

The occurrence of resonant bonding effects between spheres as well as the energetic preference for the arrangement of crystal layers into the sequence $ABAB...$, may add to the current methods for engineering 3D photonic crystals out of assemblies of colloidal spheres in liquid solutions. Experiments on gold nanospheres (radius $r \sim 10$ nm) by Kimura have shown fast formation of aggregates upon illumination at the surface-plasmon frequency ({\it i.e.} the electric-dipole mode $a_1^1$), in agreement with calculations of the photo-induced Van der Waals dipole-dipole interaction\cite{Kimura}. Burns, Fournier and Golovchenko on the other hand have concentrated on dielectric microspheres ($r \sim 700$ nm)\cite{Golov}. They have experimentally demonstrated the presence of long-range attractive forces, induced by laser light, between two polystyrene spheres in water. They argue that these are magnetic-dipole interactions, owing their existence to induced currents on the spheres. They calculate the interaction energy between two magnetic dipoles based on a simplified model that works in the long-distance limit $R \gg \lambda$, where $R$ the distance between the dipoles and $\lambda$ the wavelength of light.
From the perspective of Mie resonances this approach is justified because although the spheres Burns {\it et al} use possess many different resonances at or near the frequency of their laser light, the large inter-sphere distance ensures that the dipole approximation works to leading order. Also the electric-dipole solutions are of secondary importance since they are damped compared to the magnetic dipoles.

In contrast to the work of Kimura (metallic spheres) and that of Burns {\it et al} ($R \gg \lambda$), we are dealing with {\it dielectric} spheres at {\it close proximity} ($R \sim \lambda$). In general, the sphere-sphere interactions in our system display a {\it mixing} of EM-multipole behaviour, unless a specific condition for an isolated Mie resonance is met. For example two spheres of $r=349$ nm interact at $\sim$ 0.65 eV as oscillating magnetic dipoles, at $\sim$ 0.80 eV as oscillating electric dipoles and so on. Although the model of Burns, Fournier and Golovchenko does not apply to our system (due to the close proximity between the spheres), we believe that their physical explanation for the spatially oscillatory shear forces is basically correct: the oscillation is due to the changes in the phase shifts associated with retardation between spheres of different layers.

\section{Conclusion}
\label{sec:Conclusion}

In conclusion, we have presented a general methodology for accurate and efficient computation of EM fields and forces in matter. We have applied this first to homogeneous systems (dielectric and metallic) and studied the effects of travelling and evanescent waves on the forces exerted on a slab of material. Under specific conditions, evanescent fields can produce attractive total forces which are even stronger than the Van der Waals attraction between two half-spaces.
Next, we have concentrated on a dielectric crystal of GaP spheres arranged in a SC lattice. We observe the connection between lattice effects (band gaps, Bragg scattering) and EM forces. More interestingly, however, we find large resonant forces at frequencies corresponding to the EM eigenmodes of isolated spheres. An analogy to atomic physics proves fruitful as the EM fields on the spheres arrange into bonding and anti-bonding states, causing highly attractive and repelling interactions which may become dominant over gravitational, thermal and Van der Waals forces. We believe these effects to be experimentally observable and relevant to the formation of stable crystals of colloidal nanospheres.

\acknowledgments

Many thanks to F.-J. Garcia-Vidal for valuable discussions and comments. 
I would like to acknowledge the Trustees for the Beit Fellowship for Scientific Research, as well as my parents, Yannis and Chryssoyla, for financial assistance. - M.I. Antonoyiannakis.

\vfill
\eject

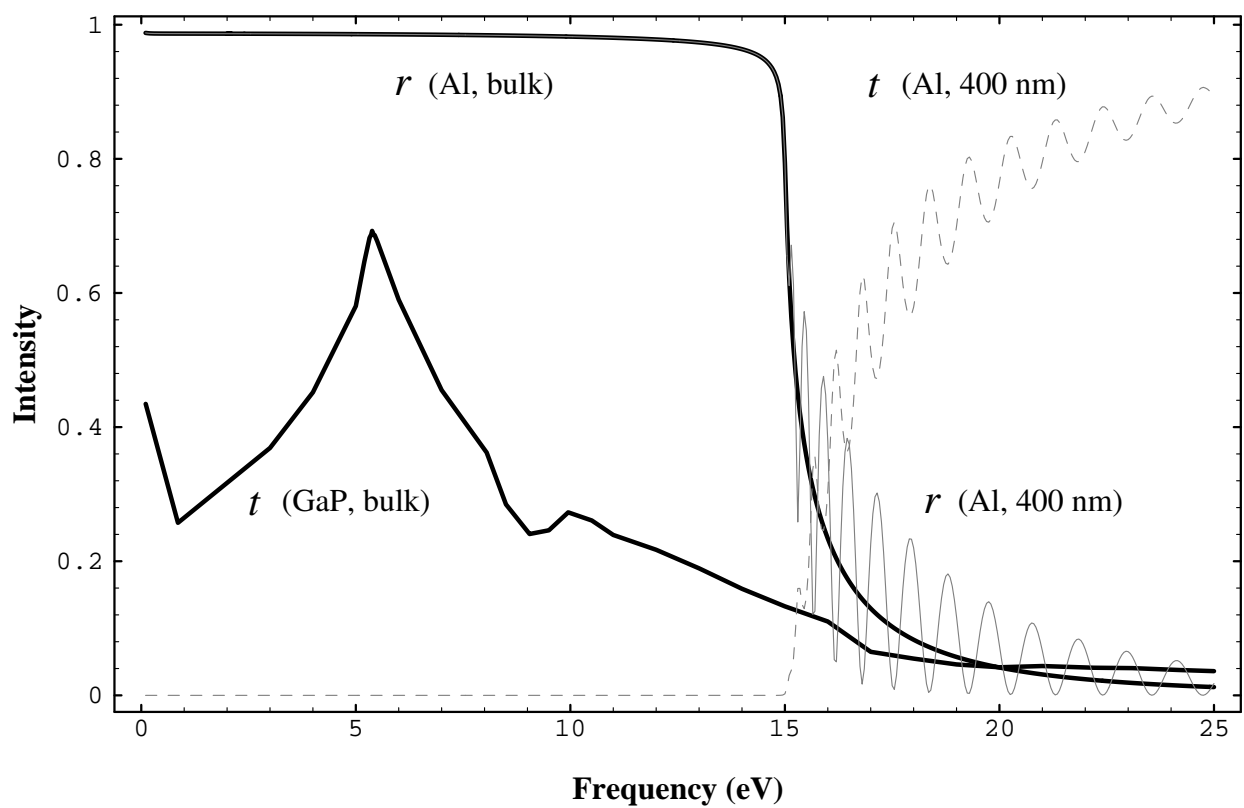
\begin{figure}
\caption{The intensity of reflected light, at normal incidence, from two semi-infinite homogeneous systems, GaP and Al, is shown in black lines. The sharp decline in reflectance for Al is due to the plasma frequency $\omega_p = 15$ eV. A finite but substantially thick (400 nm) layer of Al also reflects most light below $\omega_p$; above $\omega_p$ it displays a ripple structure due to multiple reflections of the waves from the sides of the layer (Fabry-Perot oscillations). Reflectance (transmittance) for the finite layer is shown in solid (broken) grey lines.}
\label{fig.1}
\end{figure}
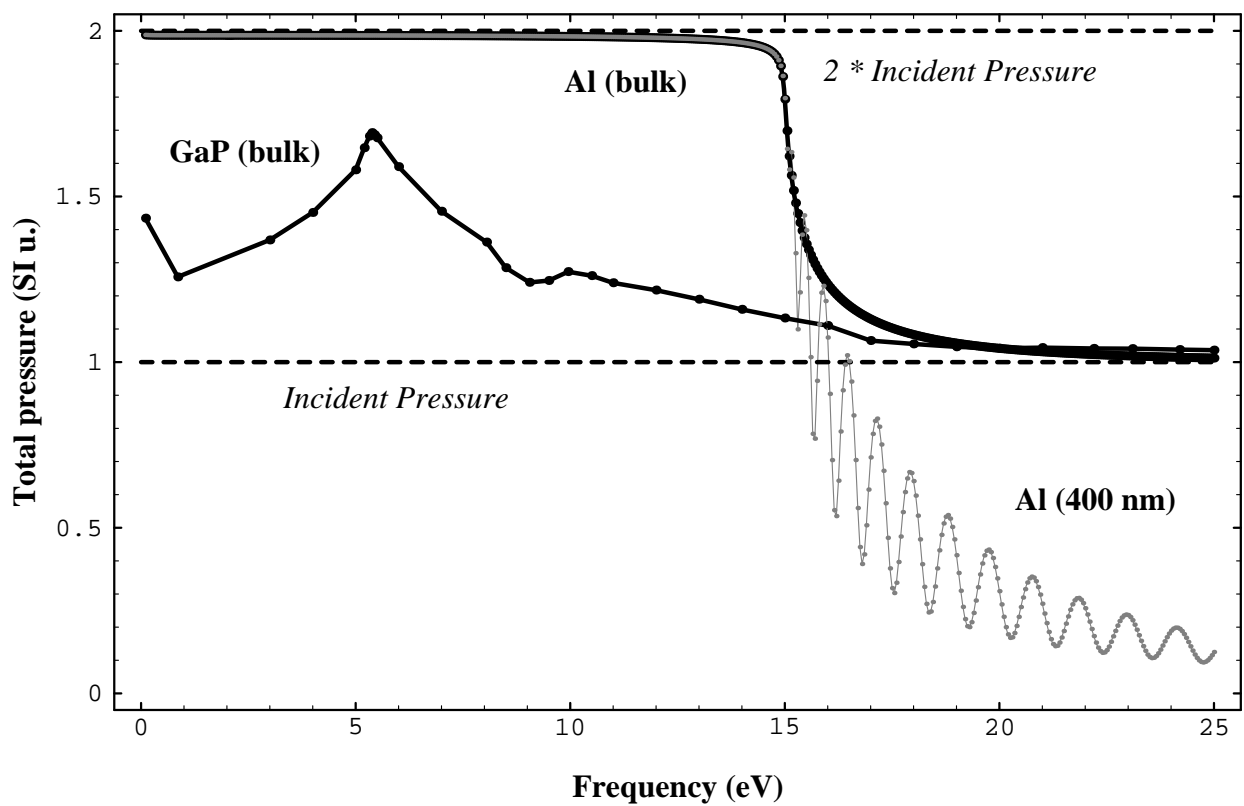
\begin{figure}
\caption{
Pressure on a homogeneous medium of infinite (GaP and Al half-spaces, black curves) and finite (Al 400 nm, grey curves) thickness. Light is normally incident along the $+z$ direction with intensity $I_0 \sim 3 \times 10^8$ W/m$^2$. Positive values for the pressure imply that the light is pushing the system. (i) For the half-spaces, the pressure is calculated with the stress-tensor method analytically (solid black lines) and numerically  (dotted black lines) in perfect agreement. It lies within $P_{inc}$ (full absorption) and $2 P_{inc}$ (full reflection) irrespective of the different dielectric properties of the GaP and Al media. (ii) In contrast, a layer of finite thickness experiences a pressure with a zero lower bound, corresponding to full transmission through the layer. For this system the pressure is calculated numerically with the stress-tensor method (solid grey line) and with the energy-gradient method (dotted grey line).}
\label{fig.2}
\end{figure}
\begin{figure}
\caption{A wave incident at angle $\theta_1$ from medium I is scattered by the interfaces with media II and III, which are separated by a thickness $d$. When $\epsilon_2 < \epsilon_1$ the fields in II become evanescent for large enough $\theta_1$.}
\label{fig.homog.layer}
\end{figure}
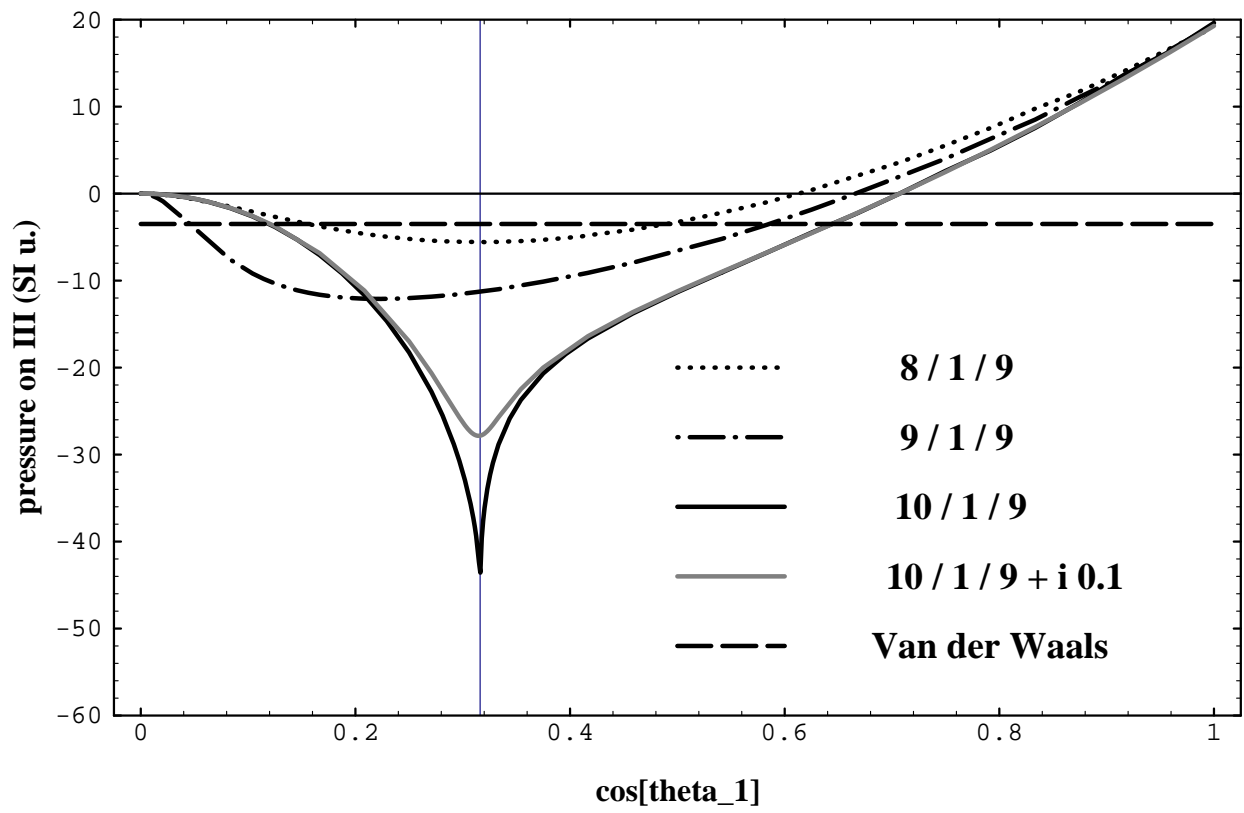
\begin{figure}
\caption{Total pressure on medium III (normal force per unit area), plotted against the angle of incidence, for (i) $\epsilon'_1 = 8, \epsilon'_2 = 1, \epsilon'_3 = 9$, (ii) $\epsilon'_1=9, \epsilon'_2 = 1, \epsilon'_3 = 9$ and (iii) $\epsilon'_1 =10, \epsilon'_2 = 1, \epsilon'_3 = 9$. For the latter system nonzero absorption ($\epsilon''_3=0.1$) is considered also (grey line). Negative pressure implies attractive force on III. The sharp attraction peak for (iii) occurs when $|\gamma_3| = 0$. The zero pressure corresponds to 
$\gamma''_2 = |\gamma_3|$. All curves are for $d=100$ nm, $\lambda/d = 100$ and an incident beam intensity $I_0=3.5 \times \,10^9$ W/m$^2$. For this system the Van der Waals attraction is smaller than the photo-induced force for most angles; however for light intensities weaker by a factor of 20 or more, or for separations smaller than 50 nm the Van der Waals force dominates all photo-induced effects over all angles.}
\label{fig.evan}
\end{figure}
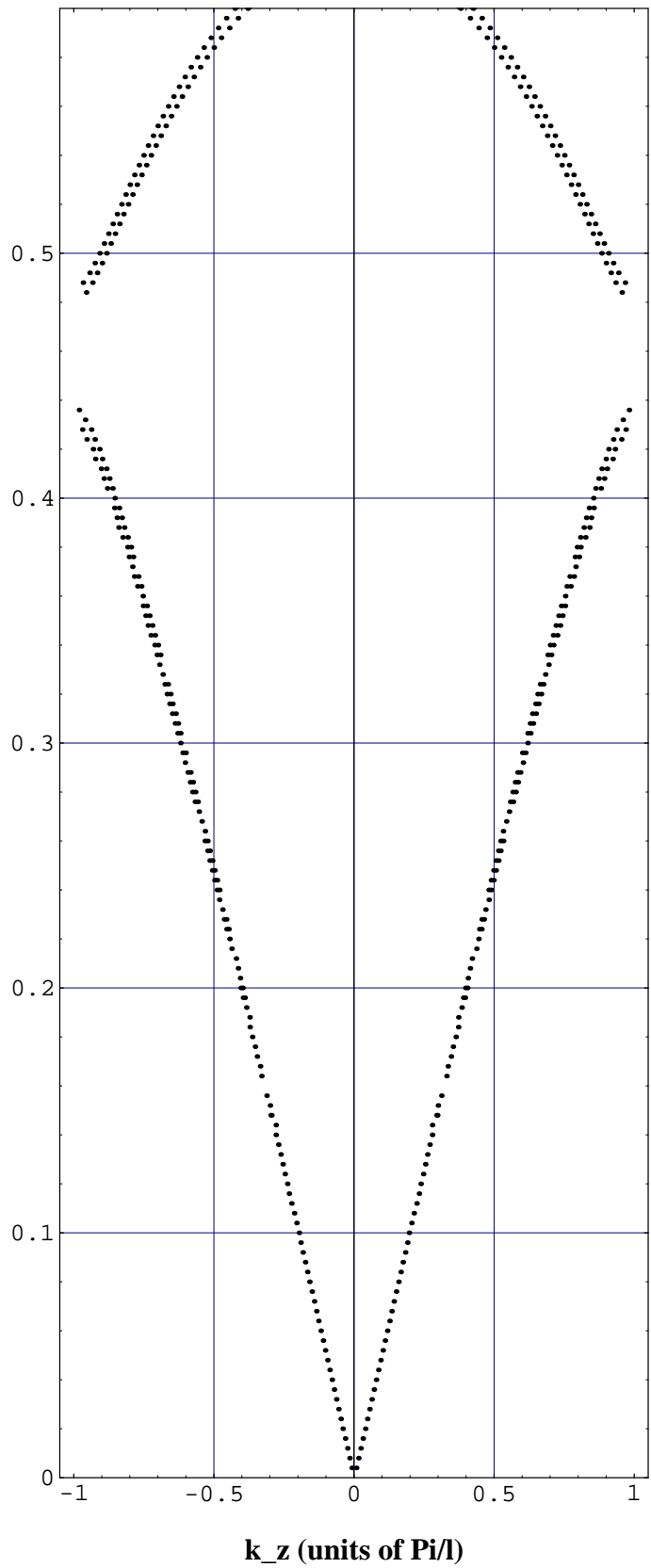
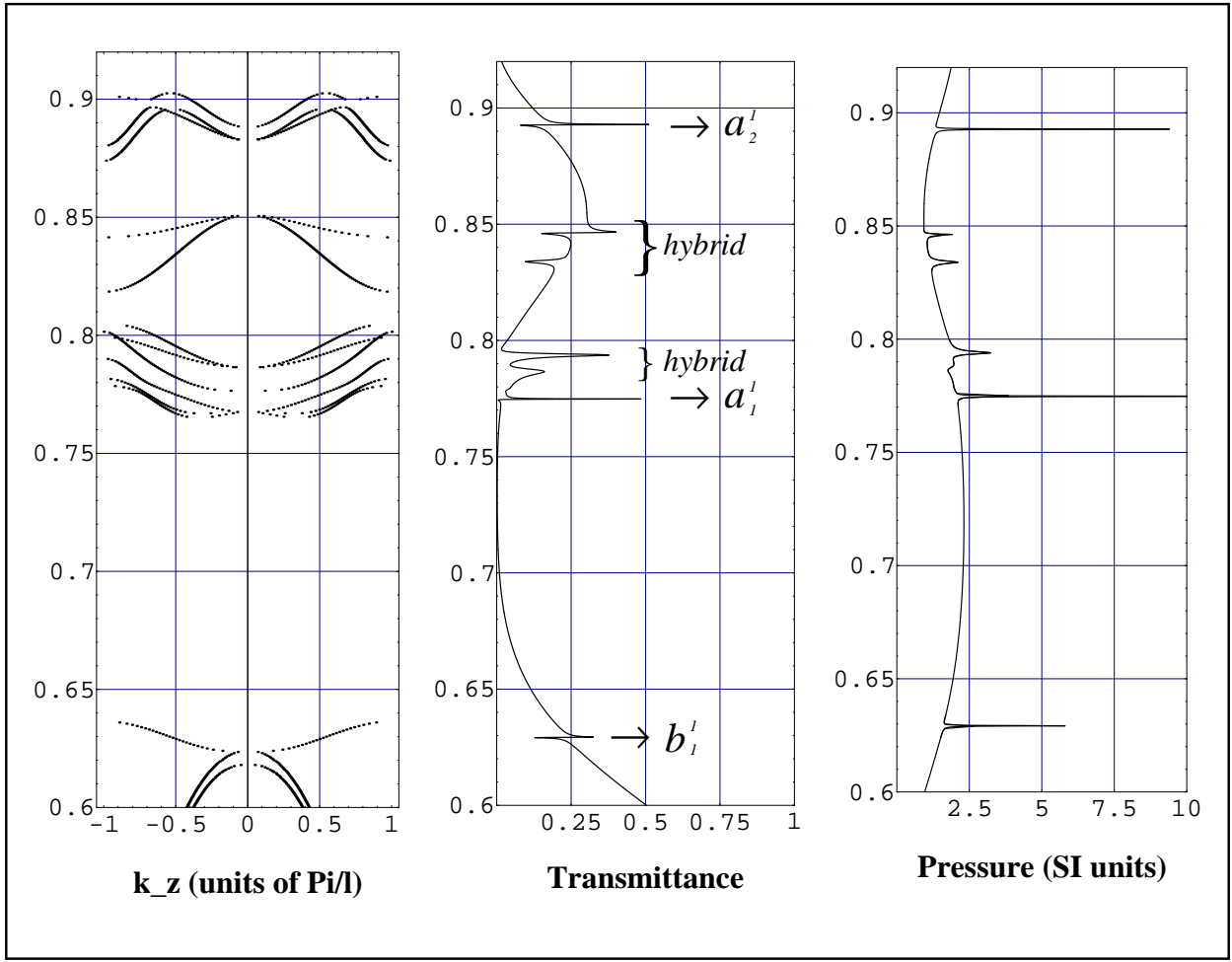
\begin{figure}
\caption{
(a,b) Photonic band structure along the $\Gamma X$ direction for a simple cubic lattice of GaP spheres ($r=365$ nm, $\epsilon=8.9$) in air. The vertical axis is in eV units for the frequency. (c) Transmittance for a crystal one-unit-cell thick and for normal incidence along ${\bf {\hat z}}$. The sharp peaks correspond in frequency to the low-dispersion modes of the band structure in (b), and occur when the incident light excites one of the EM eigenmodes of isolated spheres (Mie resonances). Some peaks can be attributed to a single resonance ($b_1^1$: magnetic dipole mode; $a_1^1$: electric dipole mode; $a_2^1$: electric quadrupole mode) while others involve mixing from more than one Mie modes (hybrids) owing to the organisation of the spheres in a lattice. (d) Normal force on one unit cell for the system of (c) and beam intensity $I_0=3 \times \,10^8$ W/m$^2$. The force scales with $I_0$.
}
\label{fig.pbs.T.force}
\end{figure}
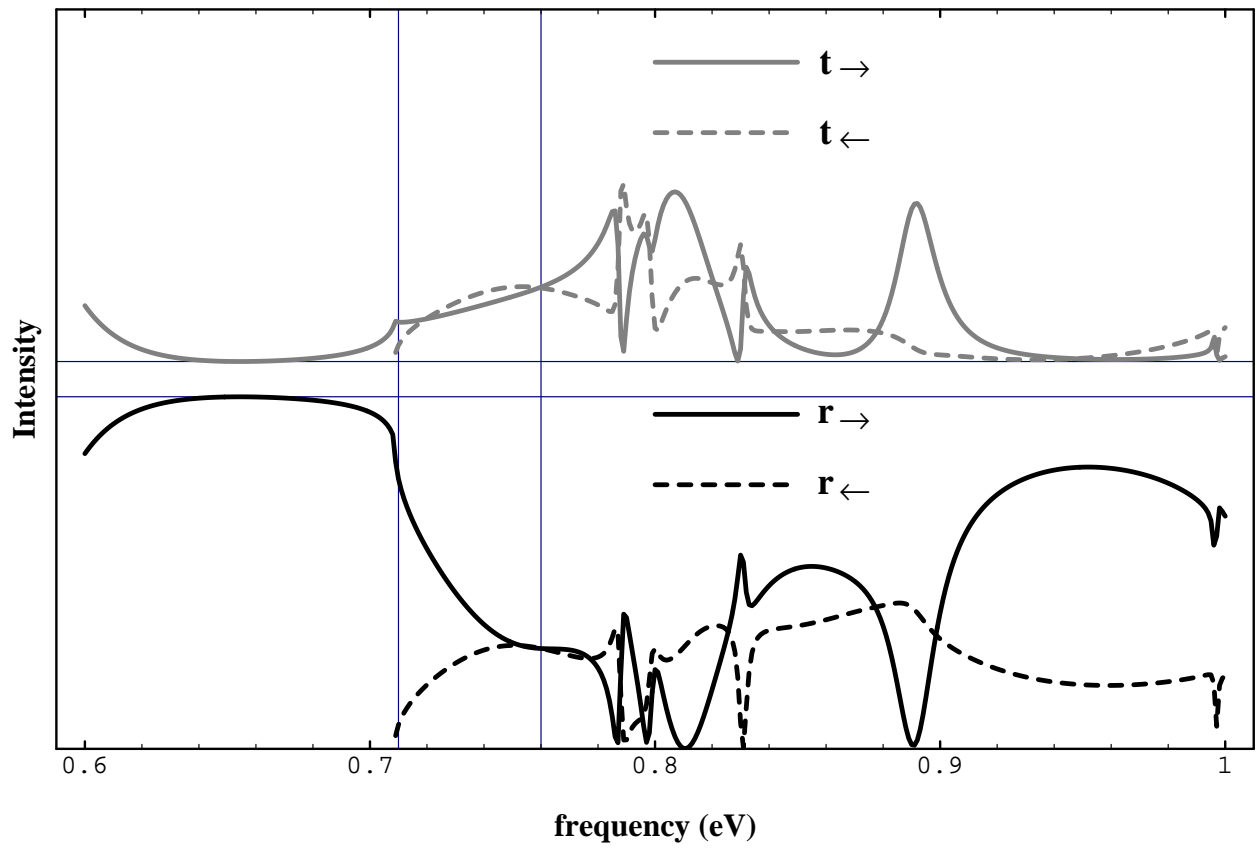
\begin{figure}
\caption{
Intensity of the specularly reflected (solid black line), directly transmitted (solid grey line), off-diagonally reflected (dashed black line) and off-diagonally transmitted (dashed grey line) beams. The angle of incidence is $\theta=70 \deg$. For low frequencies,  corresponding to wavevectors away from the edge of Brillouin zone, only the diagonal beams exist. However at the condition of Eq.\ (\ref{eq:ODS}) off-diagonal scattering starts to occur (at 0.71 eV). For the frequency range shown here only two off-diagonal beams exist. The spheres ($\epsilon=8.9$) have radius 365 nm and are suspended in air.}
\label{fig.ODS.R}
\end{figure}
\begin{figure}
\caption{Geometry of the scattered beams for off-diagonal scattering. The off-diagonal beams have negative momentum along ${\bf {\hat x}}$. At 0.76 eV, when the intensities of the waves travelling to the right and to the left balance each other (see Fig.\ \ref{fig.ODS.R}), the total scattered $x$-momentum is zero; the crystal absorbs all incident momentum along ${\bf {\hat x}}$.}
\label{fig.ODS.beams}
\end{figure}
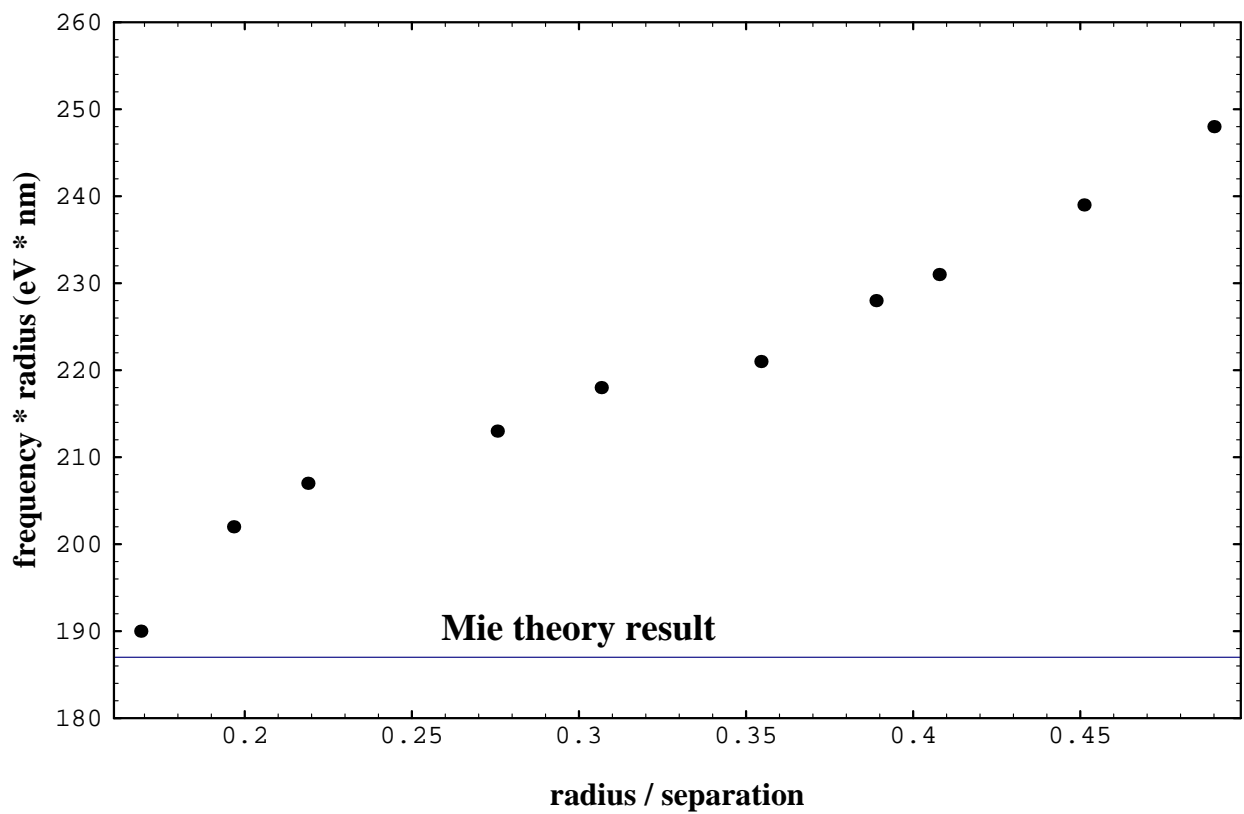
\begin{figure}
\caption{For a given Mie mode (at frequency $\omega$) of an isolated sphere (radius $r$), the  order parametre $\omega \cdot r$ is independent of the sphere size. When the spheres are put together in a lattice, interactions shift the resonances. When the radius is small compared to the separation ({\it i.e.} the unit-cell length $l$), the one-sphere result is recovered. Results here are for the $b_1^1$ mode.}
\label{fig.w.r}
\end{figure}
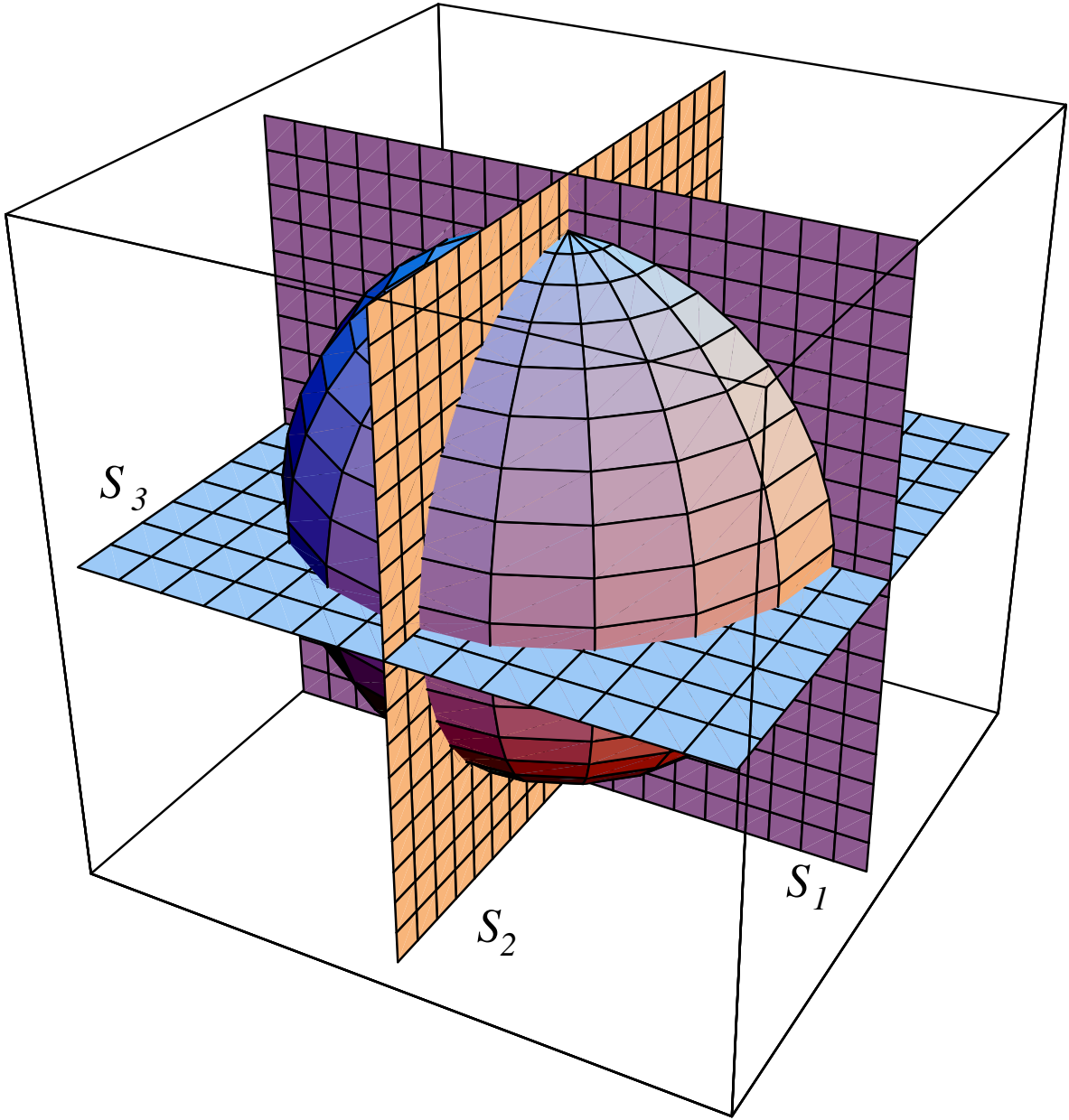
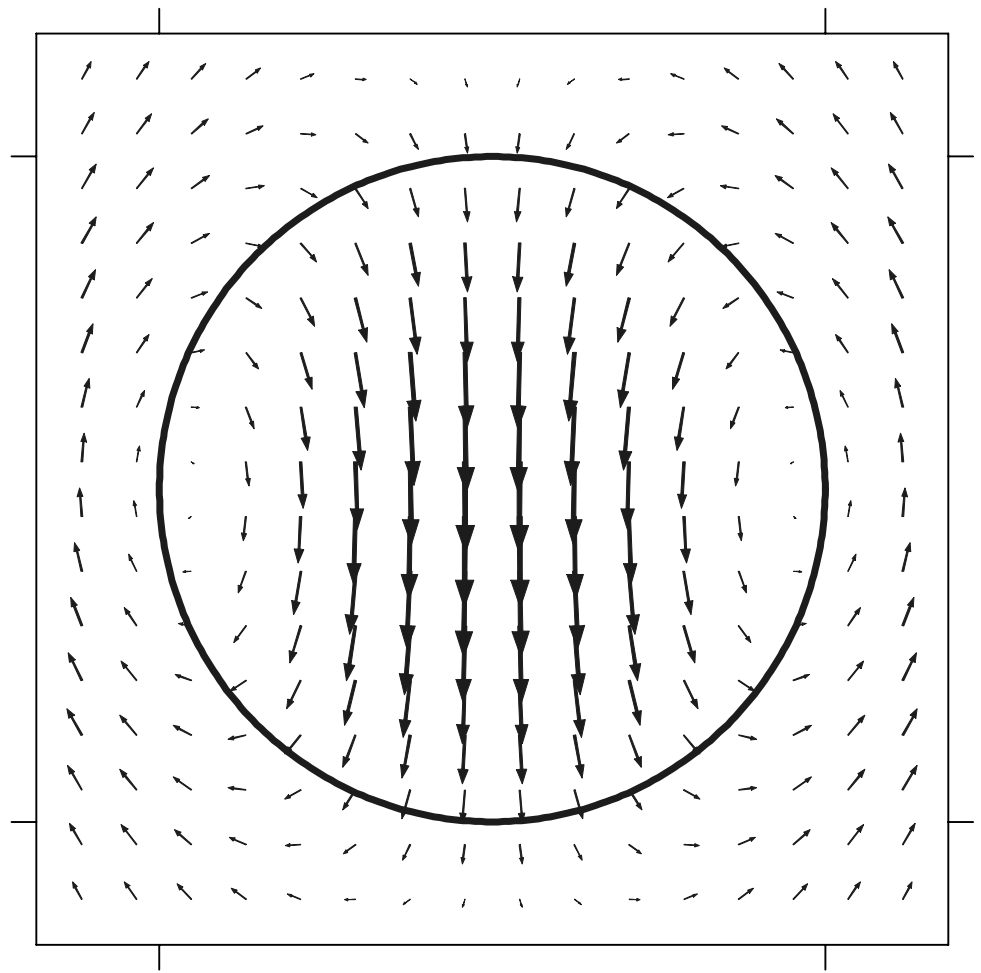
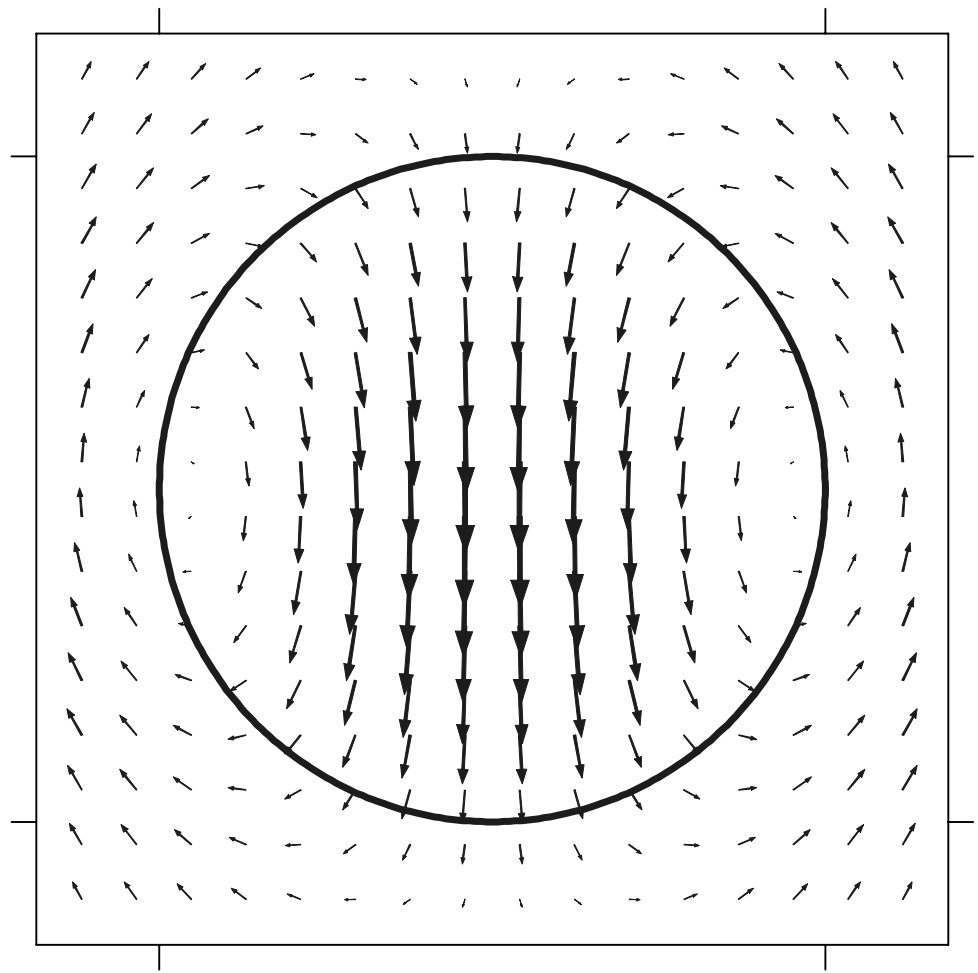
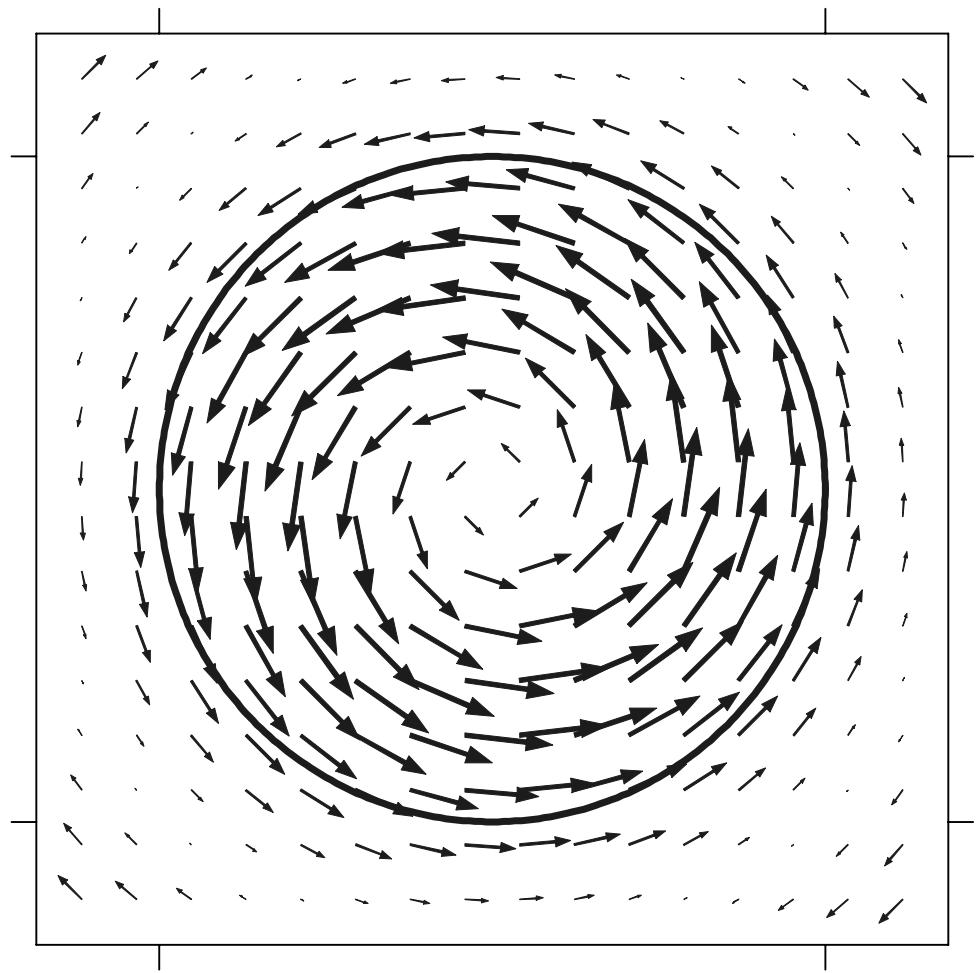
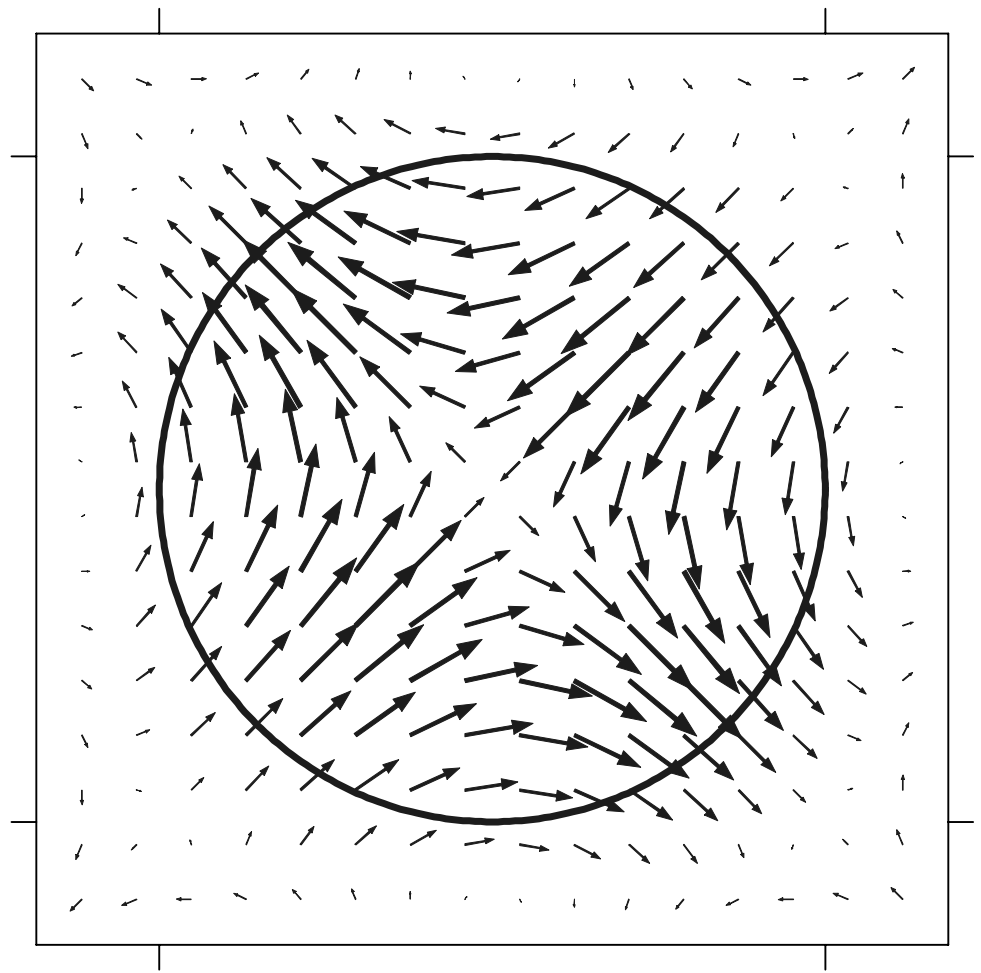
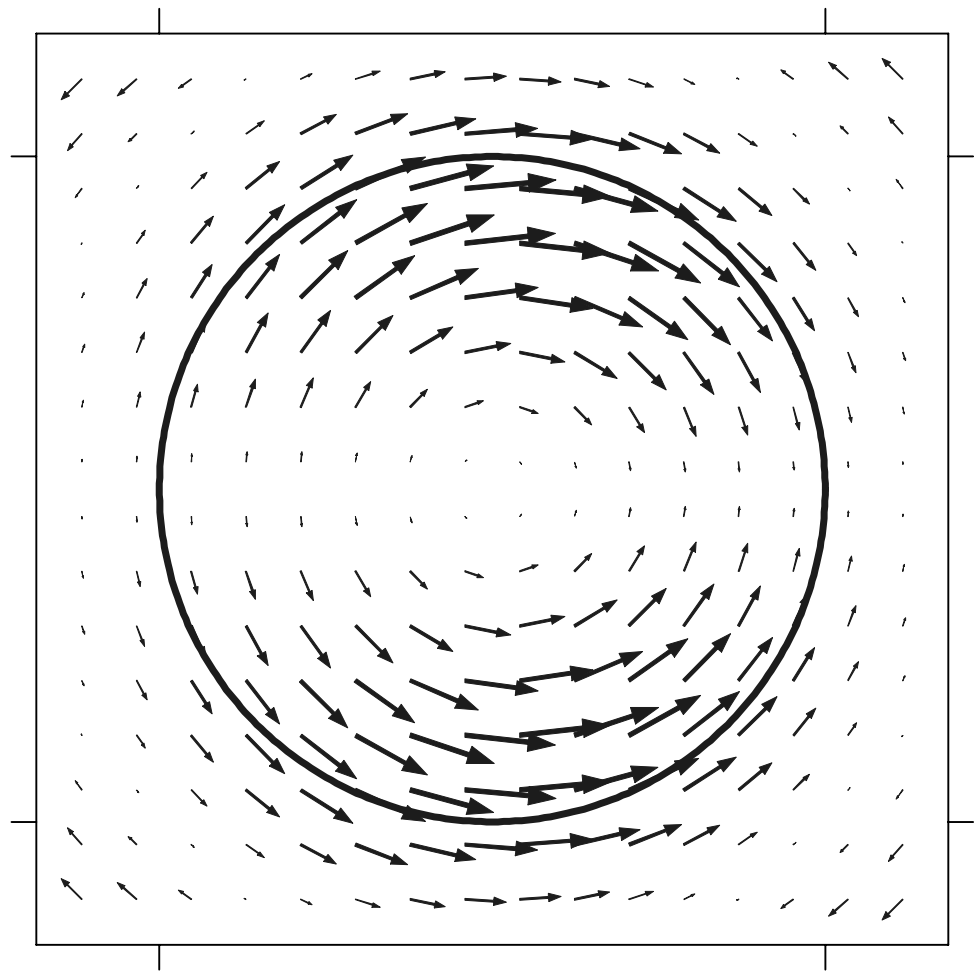
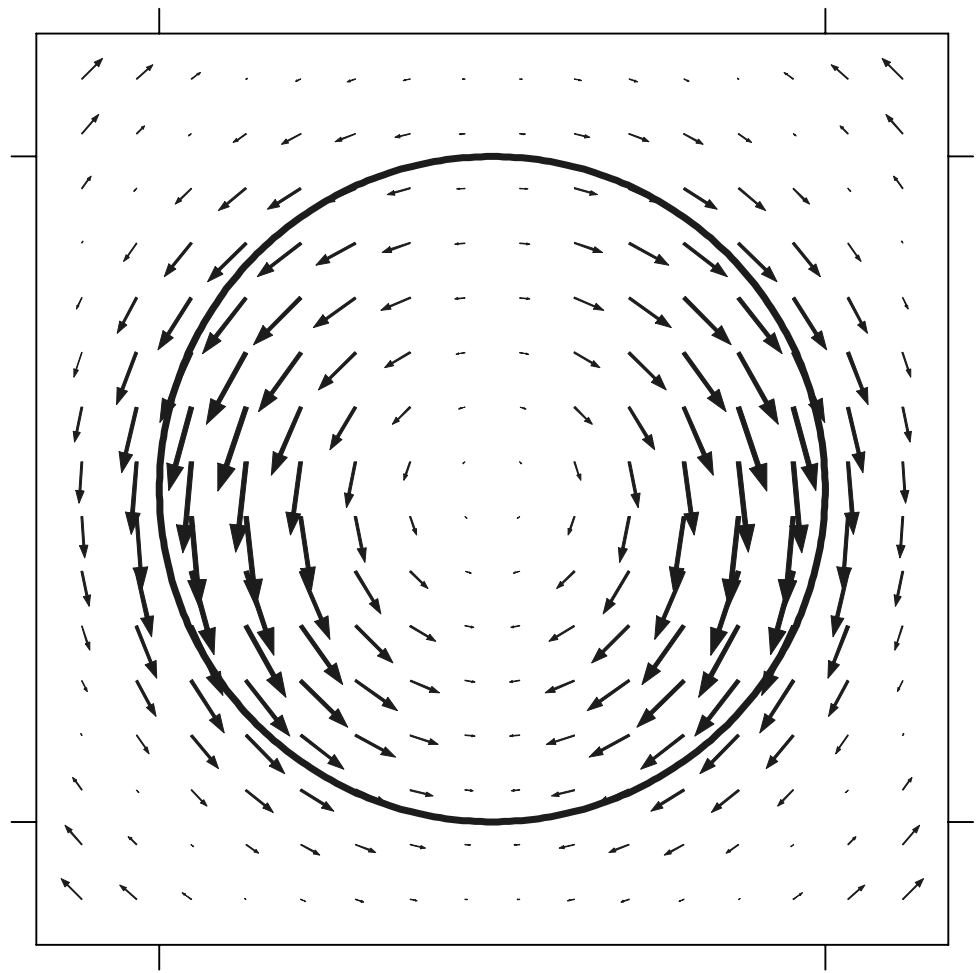
\begin{figure}
\caption{
(a): The spherical sections $S_1,S_2,S_3$ on which the field distributions are plotted. 
(b-d): Dipole mode. For the $b_1^1$ resonance (magnetic mode) the ${\bf H}$ field is plotted on the spherical sections $S_1, S_2$, while the ${\bf E}$ field is shown on $S_3$. For the $a_1^1$ electric mode the role of the ${\bf H}$ and ${\bf E}$ fields is interchanged.
(e-g): Quadrupole mode. For the $b_2^1$ resonance (magnetic mode) the ${\bf H}$ field is plotted on the spherical section $S_1$, while the ${\bf E}$ field is shown on $S_2, S_3$. For the $a_2^1$ electric mode the role of the ${\bf H}$ and ${\bf E}$ fields is interchanged.
}
\label{fig.modes}
\end{figure}
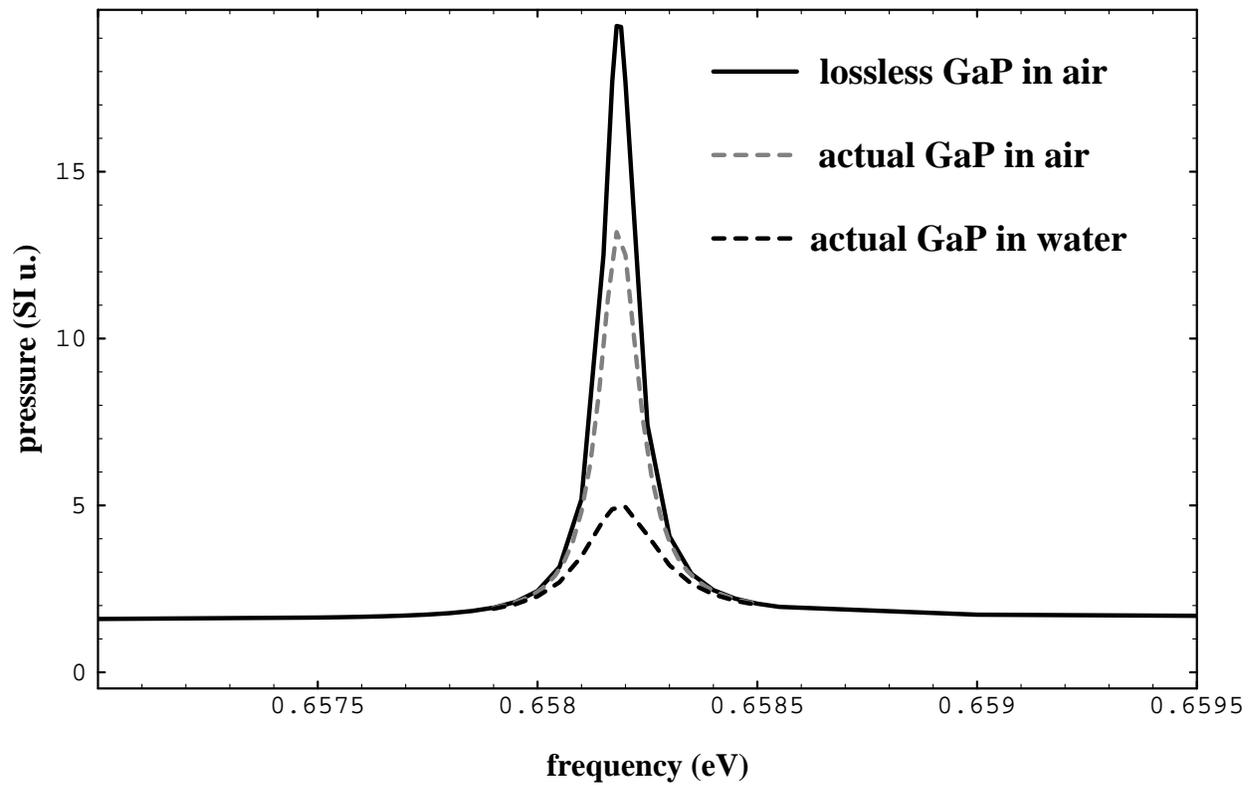
\begin{figure}
\caption{Dependence of the force on absorption. Away from resonances the forces do not suffer a sizeable change with moderate absorption. The  resonances are  sensitive to losses, but the actual amount of absorption present in GaP spheres and in an experimentally suitable liquid medium (in this case water) is small enough so the resonance effects can be clearly seen. Here we plot the normal pressure on a crystal one layer thick (GaP spheres, radius 337 nm) in air (dashed grey line) and in water (dashed black line) when a light beam, of intensity $I_0=3 \times \,10^8$ W/m$^2$, is incident along ${\bf{\hat z}}$. For comparison, the force on a hypothetical loss-free GaP sphere in air is also shown (solid black line); it is clear that GaP is a material with very low absorption in this frequency range and is therefore highly suitable for displaying the effects of resonant scattering.} 
\label{fig.f_vs_abs}
\end{figure}
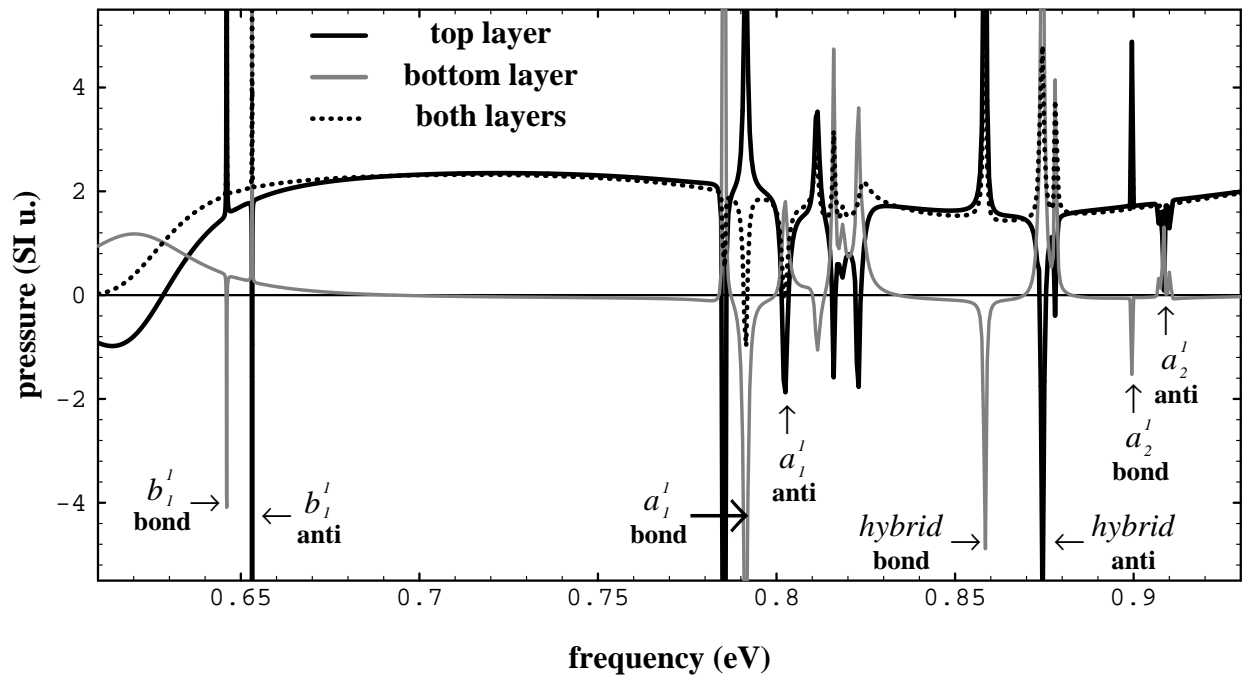
\begin{figure}
\caption{Normal pressure on a two-layer crystal of GaP spheres ($\epsilon=8.9$, radius $r=349$ nm) in air when light is incident along ${\bf{\hat z}}$ with intensity $I_0=3 \times \,10^8$ W/m$^2$. The pressure on the top (bottom) layer is shown in solid black (grey) line; on both layers ({\it i.e.} total pressure) it is shown in a dotted line. At resonant frequencies the total pressure displays peaks of mostly positive (repelling) pressure. However the resonant forces on either layer are generally opposite in direction, peak more sharply than the total force and their sign alternates with frequency, exhibiting a series of bonding/anti-bonding interactions ({\it i.e.} layers attract/repel each other). At the bonding modes the spheres of the two layers behave as oscillating EM multipoles oriented parallel to each other, at the anti-bonding modes they are anti-parallel. Various modes are seen: magnetic dipole ($b_1^1$), electric dipole ($a_1^1$), electric quadrupole ($a_2^1$) and hybrids. Compare with the force on one layer of Fig.\ \ref{fig.pbs.T.force}.}
\label{fig.f_J=1}
\end{figure}
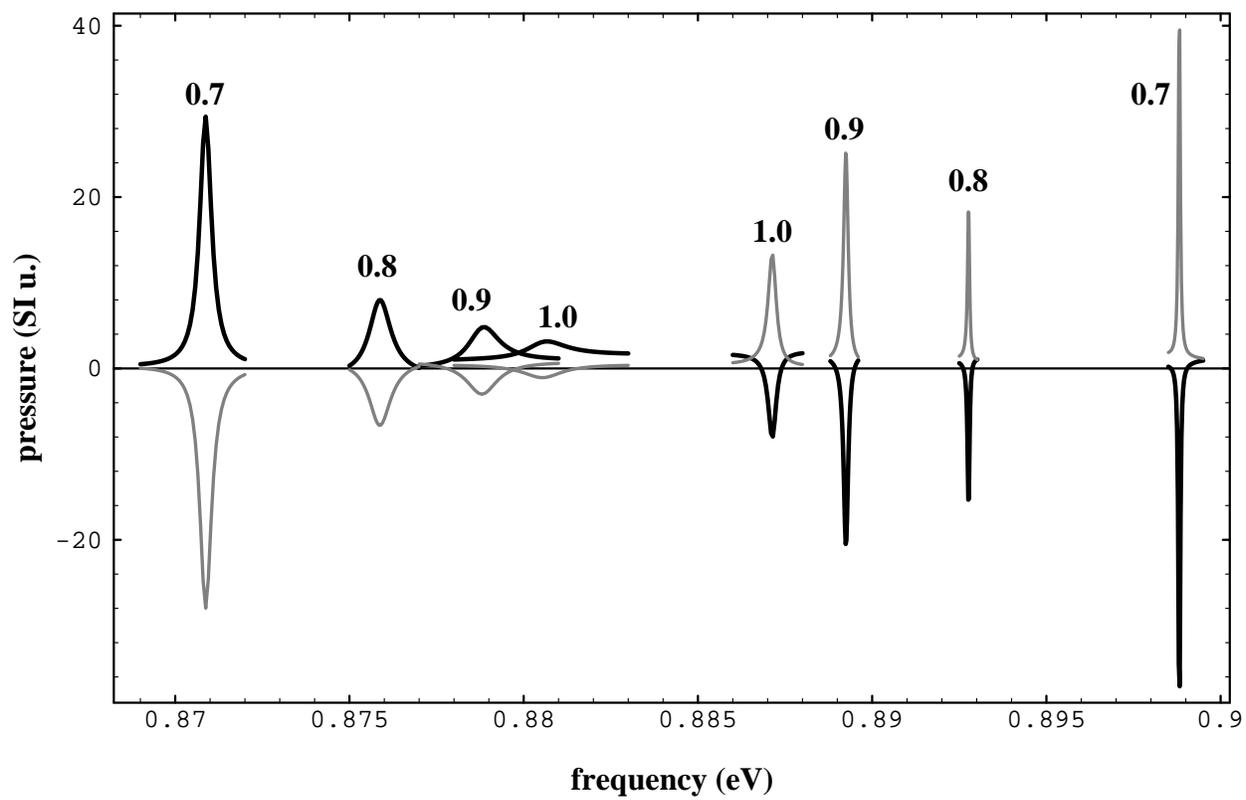
\begin{figure}
\caption{Resonant forces on a two-layer sample when light is incident normally along ${\bf{\hat z}}$ with intensity $I_0=3 \times \,10^8$ W/m$^2$ and populates the magnetic dipole mode $b_1^1$. As the two crystal layers approach each other the bonding/anti-bonding energy levels of the Mie modes on the spheres split further, and the potential well (barrier) for attraction (repulsion) between the layers becomes deeper (higher). The pressure on the top (bottom) layer is shown in black (grey). Attraction occurs when the pressure on the top layer is positive and that on the bottom layer is negative. The numbers denote the distance between two vertically adjacent spheres in multiples of the unit-cell length, $l=900$ nm. The GaP spheres have radius $r=365$ nm and are in water. The absorption of both the spheres and the medium is taken into account.} 
\label{fig.f_split.with.dist}
\end{figure}
\begin{figure}
\caption{View from above of a sample two layers thick. All the spheres of the top layer (white) are displaced along the diagonal $AC$ while the bottom-layer spheres (grey) remain fixed. The position of a top-layer sphere can be traced by the variable $\delta=x/l$, $l$ being the lattice constant for the original simple cubic lattice. }
\label{fig.diag.disp}
\end{figure}
\begin{figure}
\caption{The potential landscape experienced by sphere $A_1$ while it and all other top-layer spheres are being displaced along the surface diagonal $AC$ relative to the bottom-layer spheres. The potential wells at $\delta_{init}=0$ and $\delta_{fin}=1.0$ lead to a stacking order $AAA...$, whereas the deepest well at $\delta_{mid}=0.5$ favours a sequence $ABAB...$. The  potential energy is in units of $10^{-18}$ J. Results here are for the bonding $b_1^1$ mode of spheres of radius $r=240$ nm, $\epsilon=8.9$ and a lattice constant $l=900$ nm.}
\label{fig.pot.wells}
\end{figure}
\begin{table}
\caption{Mie modes of an isolated GaP sphere ($\epsilon=8.9$) and their corresponding {\it order parametre} (mode frequency times sphere radius) in units of eV $\times$ nm. $b_n^i$ $(a_n^i)$ modes represent magnetic (electric) multipoles oscillating at the light frequency. 
\label{table1}}
\begin{tabular}{cccccccccccc}
$b_1^1$&$a_1^1$&$b_2^1$&$a_2^1$&$b_3^1$&$a_3^1$&$b_1^2$&$a_1^2$&$b_4^1$&$a_4^1$&$b_2^2$&$b_5^1$\\
\tableline
187&214&267&313&346&405&405&410&423&491&493&499\\
\end{tabular}
\end{table}

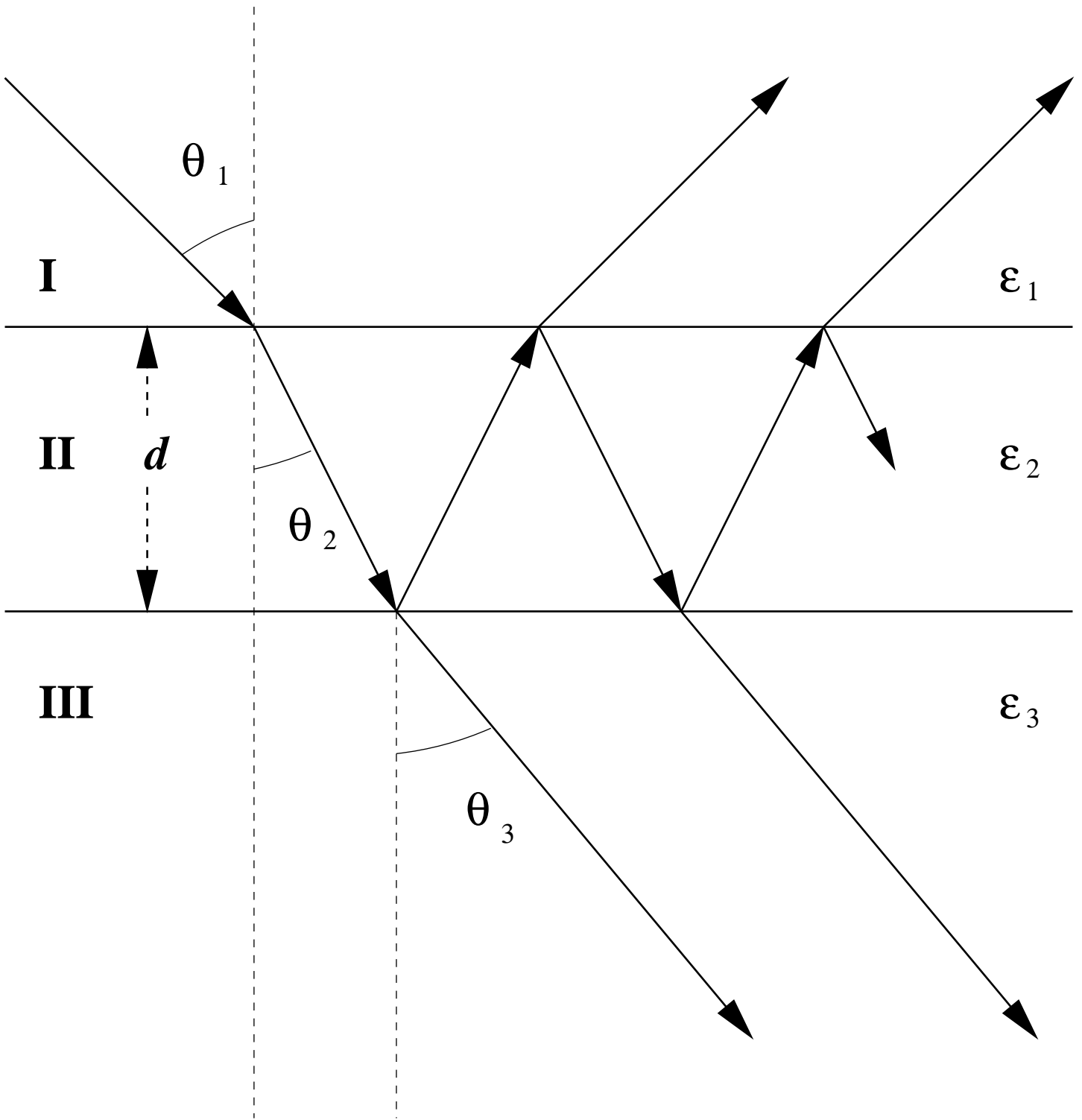
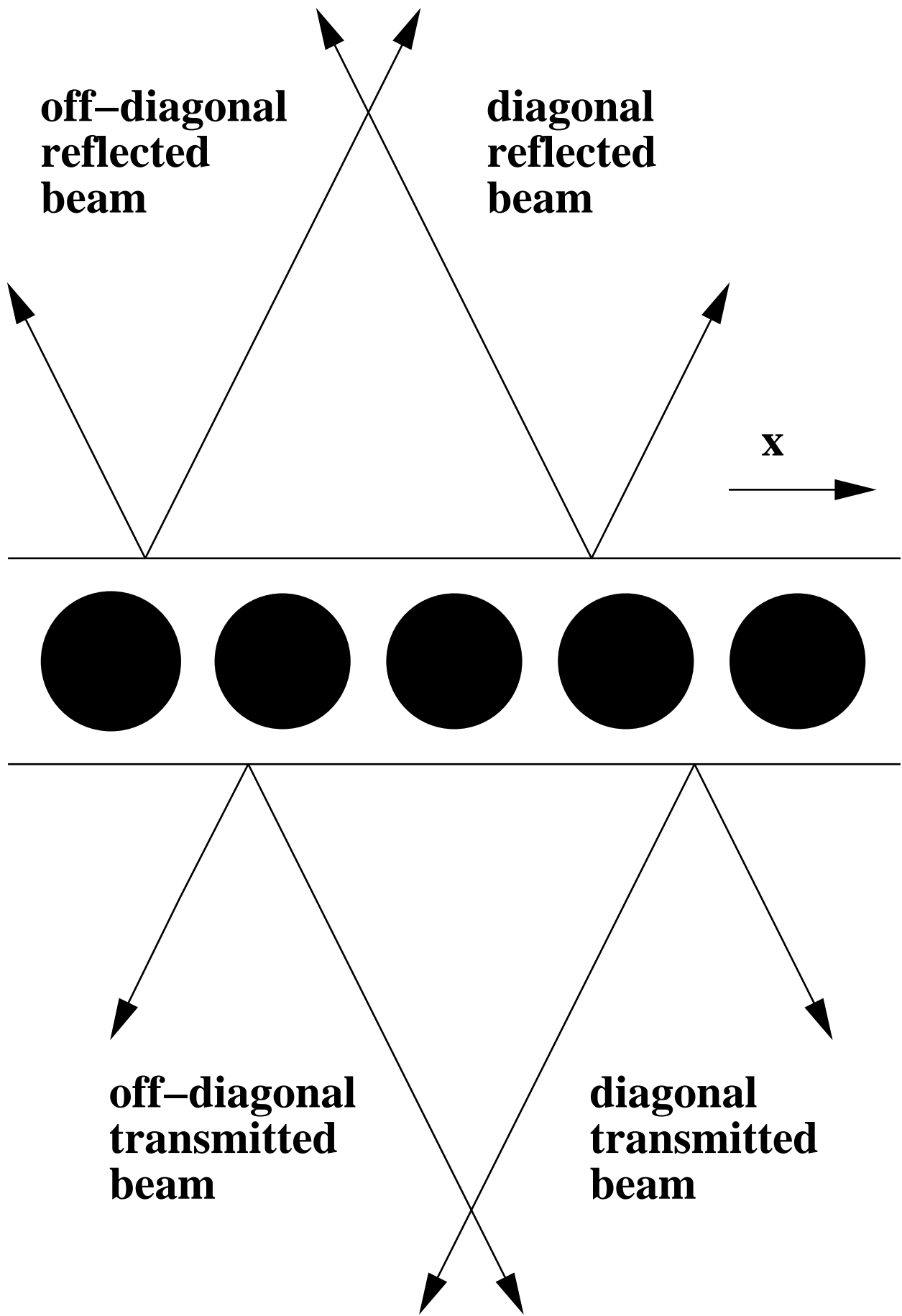
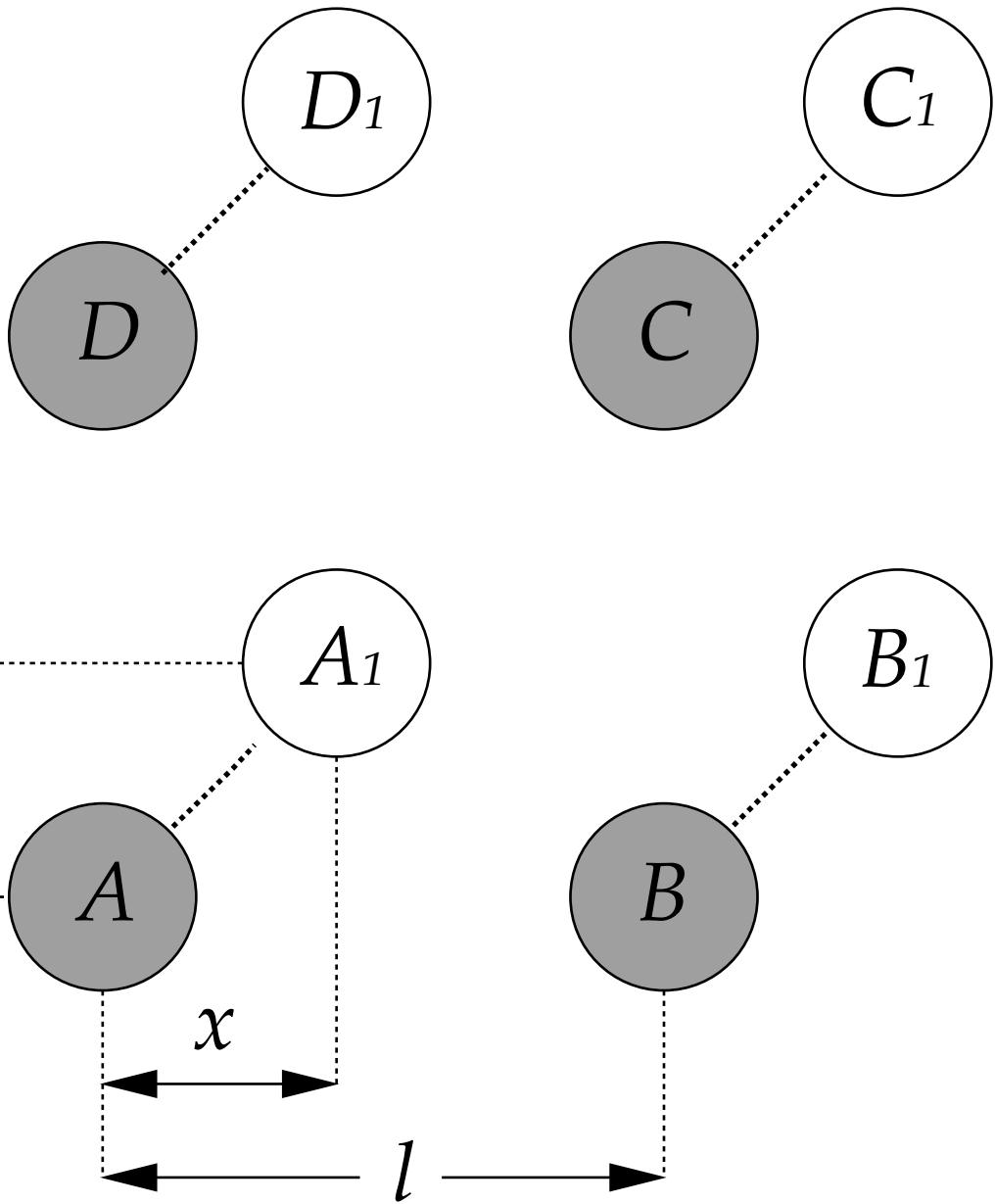
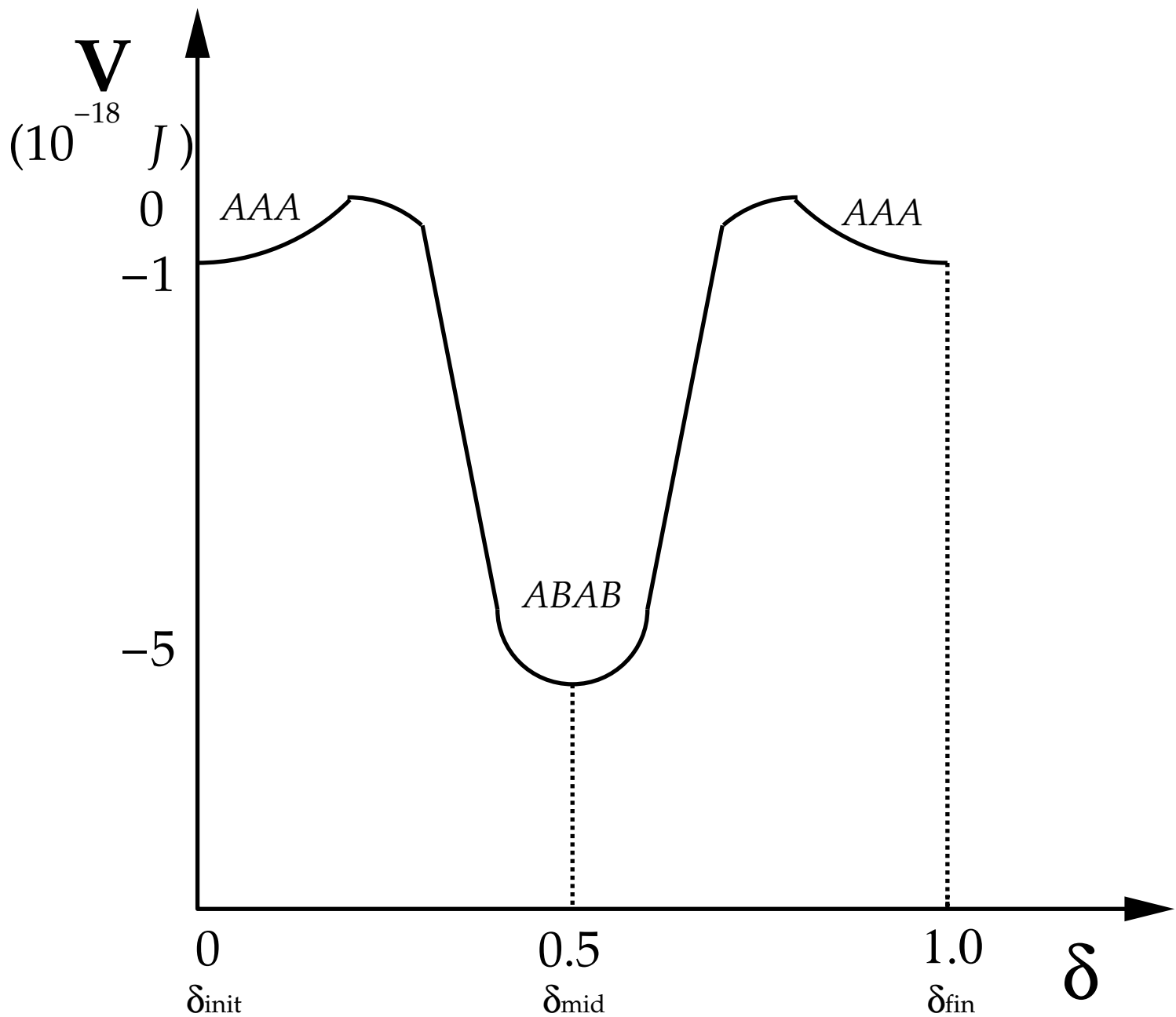
\end{document}